\documentclass[12pt]{article}
\usepackage{graphicx}
\usepackage{tabularx}
\usepackage{epstopdf}
\usepackage{epsfig}
\usepackage{amsmath}
\usepackage{amsfonts}
\usepackage{amssymb}
\usepackage{colordvi}
\usepackage{multirow}
\usepackage{subfigure}
\usepackage{rotating}
\usepackage{array}
\def\lsim{\mathrel{\raise.3ex\hbox{$<$\kern-.75em\lower 1ex\hbox{$\sim$}}}}
\def\gsim{\mathrel{\raise.3ex\hbox{$>$\kern-.75em\lower 1ex\hbox{$\sim$}}}}

\def\be{\begin{equation}}
\def\ee{\end{equation}}
\def\bea{\begin{eqnarray}}
\def\eea{\end{eqnarray}}

\begin{document}

\title{Nonstandard neutrino interactions at \\DUNE, T2HK and T2HKK }

\author{Jiajun Liao,$^1$ Danny Marfatia,$^1$ and Kerry Whisnant$^2$\\
\\
\small\it $^1$Department of Physics and Astronomy, University of Hawaii-Manoa, Honolulu, HI 96822, USA\\
\small\it $^2$Department of Physics and Astronomy, Iowa State University, Ames, IA 50011, USA}
\date{}

\maketitle

\begin{abstract}

We study the matter effect caused by nonstandard neutrino interactions
(NSI) in the next generation long-baseline neutrino experiments, DUNE, T2HK
and T2HKK. If multiple NSI parameters are nonzero, the potential of these
experiments to detect CP violation, determine the mass hierarchy and constrain NSI is severely impaired by degeneracies between the NSI parameters and by the generalized mass hierarchy degeneracy. In particular, a cancellation
between leading order terms in the appearance channels when
\mbox{$\epsilon_{e\tau} = \cot\theta_{23} \epsilon_{e\mu}$}, strongly
affects the sensitivities to these two NSI parameters at T2HK and
T2HKK. We
also study the dependence of the sensitivities on the true CP phase
$\delta$ and the true mass hierarchy, and find that overall DUNE has the best
sensitivity to the magnitude of the NSI parameters, while T2HKK has
the best sensitivity to CP violation whether or not there are NSI.  Furthermore,
for T2HKK a smaller off-axis angle for the Korean detector is better overall. We find that due to the structure
of the leading order terms in the appearance channel probabilities, the
NSI sensitivities in a given experiment are similar for both mass hierarchies, modulo the phase change $\delta \to \delta + 180^\circ$.

\end{abstract}

\newpage

\section{Introduction}

The success of neutrino oscillation experiments in the last few
decades is a significant triumph in modern physics, and the masses and
mixing angles of neutrinos have been incorporated into the standard
model (SM)~\cite{Agashe:2014kda}. The data from a plethora of neutrino experiments
using solar, atmospheric, reactor, and
accelerator neutrinos can be explained in the framework of three
neutrino mixing, in which the three known neutrino flavor eigenstates
($\nu_e,\nu_\mu,\nu_\tau$) are quantum superpositions of three mass
eigenstates ($\nu_1,\nu_2,\nu_3$). In the SM with three massive
neutrinos, the neutrino oscillations probabilities are determined by
six oscillation parameters: two mass-squared differences ($\delta
m_{21}^2$, $\delta m_{31}^2$), three mixing angles ($\theta_{12},
\theta_{13}, \theta_{23}$) and one Dirac CP phase
$\delta$. Currently, the first five oscillation parameters have been
well determined (up to the sign of $\delta m_{31}^2$) to the few percent level, and the main physics goals
of current and future neutrino experiments are to measure the Dirac CP
phase and to determine the neutrino mass hierarchy (MH), i.e., the sign of $\delta m_{31}^2$,
and the octant
of $\theta_{23}$, i.e., whether $\theta_{23}$ is larger or smaller than 45$^\circ$. Future neutrino oscillation experiments will
reach the sensitivity to do precision tests of the three neutrino
oscillation paradigm and probe new physics beyond the SM.

A model-independent way of studying new physics in neutrino
oscillation experiments is provided by the framework of nonstandard
interactions (NSI); for recent reviews see
Ref.~\cite{Ohlsson:2012kf}. In this framework, new
physics is parametrized as NSI at production, detection and in propagation
according to their effects on the experiments. Since
model-independent bounds on the production and detection NSI are
generally an order of magnitude stronger than the matter
NSI~\cite{Biggio:2009nt}, we neglect production and detection NSI in
this work, and focus on matter NSI, which can be described by
dimension-six four-fermion operators of the
form~\cite{Wolfenstein:1977ue}
\be
  \label{eq:NSI}
  \mathcal{L}_\text{NSI} =2\sqrt{2}G_F
   \epsilon^{\mathfrak{f}C}_{\alpha\beta} 
        \left[ \overline{\nu}_\alpha \gamma^{\rho} P_L \nu_\beta \right] 
        \left[ \bar{\mathfrak{f}} \gamma_{\rho} P_C \mathfrak{f} \right] + \text{h.c.}\,,
\ee
where $\alpha, \beta=e, \mu, \tau$, $C=L,R$, $\mathfrak{f}=u,d,e$, and
$\epsilon^{\mathfrak{f}C}_{\alpha\beta}$ are dimensionless parameters that quantify the
strength of the new interaction in units of $G_F$.

The Hamiltonian for neutrino propagation in the presence of matter NSI can be written as
\be
H = {1\over2E} \left[  U
\left( \begin{array}{ccc}
0 & 0 & 0\\ 0 & \delta m^2_{21} & 0 \\ 0 & 0 & \delta m^2_{31}
\end{array} \right)
U^\dagger + V\right]\,,
\ee
where $U$ is the Pontecorvo-Maki-Nakagawa-Sakata mixing matrix~\cite{Agashe:2014kda}
\be
U = \left(\begin{array}{ccc}
c_{13} c_{12} & c_{13} s_{12} & s_{13} e^{-i\delta}
\\
-s_{12} c_{23} - c_{12} s_{23} s_{13} e^{i\delta} &
c_{12} c_{23} - s_{12} s_{23} s_{13} e^{i\delta} &
c_{13} s_{23}
\\
s_{12} s_{23} - c_{12} c_{23} s_{13} e^{i\delta} &
-c_{12} s_{23} - s_{12} c_{23} s_{13} e^{i\delta} &
c_{13} c_{23}
\end{array} \right)\,,
\ee
and $V$ represents the potential from interactions of neutrinos
in matter,
\be
V =  A \left(\begin{array}{ccc}
1 + \epsilon_{ee} & \epsilon_{e\mu}e^{i\phi_{e\mu}} & \epsilon_{e\tau}e^{i\phi_{e\tau}}
\\
\epsilon_{e\mu}e^{-i\phi_{e\mu}} & \epsilon_{\mu\mu} & \epsilon_{\mu\tau}e^{i\phi_{\mu\tau}}
\\
\epsilon_{e\tau}e^{-i\phi_{e\tau}}& \epsilon_{\mu\tau}e^{-i\phi_{\mu\tau}} & \epsilon_{\tau\tau}
\end{array}\right)\,.
\label{eq:V}
\ee
Here $c_{jk} \equiv \cos\theta_{jk}$, $s_{jk} \equiv \sin\theta_{jk}$,
$A \equiv 2\sqrt2 G_F N_e E$, $N_e$ is the number density of
electrons, the unit contribution to $V_{ee}$ arises from the standard 
charged-current interaction. The effective NSI parameters are given by
\be
\epsilon_{\alpha\alpha} =
\sum_{\mathfrak{f},C} \epsilon^{\mathfrak{f},C}_{\alpha\alpha}
{N_{\mathfrak{f}}\over N_e} \quad{\rm and}\quad
\epsilon_{\alpha\beta}e^{i\phi_{\alpha\beta}} =
\sum_{\mathfrak{f},C} \epsilon^{\mathfrak{f},C}_{\alpha\beta}\,
{N_{\mathfrak{f}}\over N_e}\,\, (\alpha\ne\beta)\,,
\ee
where $N_{\mathfrak{f}}$ is the number density for fermion $\mathfrak{f}$. In the earth,
$N_u \simeq N_d \simeq 3N_e$.
The diagonal terms in $V$ are real, and since the neutrino oscillation
probabilities are not affected by a subtraction of a term
proportional to the identity matrix, one of the diagonal terms can
be chosen to be 0.  The off-diagonal terms are in general complex.

Since neutral-current interactions affect neutrino propagation
coherently, long-baseline neutrino experiments with a well-understood
beam and trajectory are an ideal place to probe matter
NSI. Studies of matter NSI effects in the MINOS experiment have been
performed in Ref.~\cite{Blennow:2007pu} and by the MINOS
collaboration~\cite{Adamson:2013ovz}. NSI analyses related to the
currently running T2K~\cite{Abe:2011ks} and
NO$\nu$A~\cite{Ayres:2004js} experiments can be found in
Refs.~\cite{coelho,Liao:2016hsa}. However, due to large systematic uncertainties and
limited statistics, these experiments cannot make a definitive
measurement of the matter NSI.

The next generation long-baseline neutrino experiments,
DUNE (Deep Underground Neutrino Experiment)~\cite{Acciarri:2015uup}, T2HK (Tokai-to-Hyper-Kamiokande)~\cite{HK} and T2HKK (Toaki-to-Hyper-Kamiokande-and-Korea)~\cite{T2HKK}
will collect much more data than the current
experiments. With improved systematic uncertainties, these next
generation experiments will reach the sensitivity to
discover NSI in the neutrino
sector.  Studies of matter NSI at DUNE and T2HK can be found in
Refs.~\cite{Liao:2016hsa, dunensi, Coloma:2015kiu,  Agarwalla:2016fkh}. Studies of NSI
with a second detector in Korea, in addition to the Kamiokande
detector, can be found in Ref.~\cite{Oki:2010uc}.\footnote{In Refs.~\cite{dunensi, Coloma:2015kiu,Oki:2010uc}, the mass hierarchy is assumed to be known. As noted in Ref.~\cite{Liao:2016hsa}, and explained in Ref.~\cite{Coloma:2016gei}, if $\epsilon_{ee}$ is $\cal O$(1), the mass hierarchy can not be determined at long-baseline experiments, which in turn strongly affects a determination of the CP phase. } 

In this paper, we use the new detector
configuration proposed in the Hyper-Kamiokande proposal~\cite{HK} and the
fluxes~\cite{fluxes} provided by the Hyper-K collaboration to study the performance of
the T2HK and T2HKK experiments in the presence of NSI, and compare
their sensitivities with the sensitivity of the DUNE experiment.
 In Section~2, we describe the
experiments considered in this work. In Section~3, we discuss the
sensitivities to SM and NSI parameters in each experiment. We summarize our
results in Section~4.

\section{Experiments}
\subsection{Experimental configurations}
The main features of the three next generation long-baseline neutrino experiments we consider are summarized in Table~\ref{tb:exp} and details are described below.
\begin{description}
 \item[DUNE:] The DUNE experiment sends neutrinos from Fermilab to the
   Homestake mine in South Dakota with a baseline of 1300~km. We
   followed the DUNE CDR~\cite{Acciarri:2015uup} that uses a
   40~kton liquid argon (LAr) detector sitting on axis with respect to
   the beam direction. There is a range of beam design options and
   here we choose the optimized design, which provides a better
   sensitivity in the appearance channel than the reference
   design. The optimized design utilizes an 80~GeV proton beam with a
   power of 1.07~MW, which corresponds to $1.47\times 10^{21}$ protons
   on target (POT) per year. We assume 3.5 years of running time in
   both neutrino and antineutrino beam modes, which gives a total exposure
   of 300 kt$\cdot$MW$\cdot$years.
 \item[T2HK:] The T2HK~\cite{HK} experiment uses an upgraded 30~GeV
   J-PARC beam with a power of 1.3~MW, which corresponds to $2.7\times
   10^{21}$ POT per year. The Hyper-K detector is located 295~km away
   from the source $2.5^\circ$ off-axis so that it detects a
   narrow band beam with an unoscillated spectrum peaked at 0.6~GeV. Of the three
   detector configurations in the Hyper-K design report~\cite{HK}, 
    we choose the 2TankHK-staged configuration, which has one
   tank taking data for 6 years and a second tank is added for another 4
   years. Each tank has 40\% photocoverage and contains a water
   Cherenkov (WC) detector with 0.19 Mton fiducial mass. We assume the
   running times between neutrino and antineutrino modes have a $1:3$
   ratio.

 \item[T2HKK:] The T2HKK~\cite{T2HKK} experiment has one
   detector in the Kamioka mine, and a second detector in Korea. The
   Hyper-K detector (HK) is located in the same place as the T2HK
   experiment, with a $2.5^\circ$ off-axis-angle and a 295~km
   baseline. For the Korean detector (KD), we consider two options for
   the off-axis-angle: (a) T2HKK-2.5 with the same $2.5^\circ$
   off-axis-angle, and (b) T2HKK-1.5 with a $1.5^\circ$
   off-axis-angle. Both KD options are at a baseline of 1100~km. 
   We assume the same neutrino beam as the T2HK experiment
   with an integrated beam power of 13 MW$\cdot$years, which
   corresponds to a total 2.7$\times 10^{22}$ POT. The total running
   time is 10 years with a ratio of $1:3$ between neutrino and
   antineutrino modes.
 
\end{description}
\begin{table}
\caption{Comparison of the experiments considered in this work.}
\begin{center}
\begin{tabular}{|m{2.2cm}|c|m{5cm}|m{2.2cm}|m{2.2cm}|}
\hline
Experiment & $\dfrac{L (\text{km})}{E_\text{peak} (\text{GeV})}$ & $\nu$ + $\bar{\nu}$ Exposure\,\,\,\,\,\,\,\,\,\,\,\,  \,\,\,\,\,\,                 (kt$\cdot$MW$\cdot 10^7$s) & Signal\quad norm.\quad\quad uncertainty & Background\quad norm.\quad\quad uncertainty \\
\hline
DUNE \quad(LAr) & $\dfrac{1300}{3.0}$ & 264  + 264\,\,\,\,\, \,\,\,\,\,\,\,\,\,\,\,\,\,\,\,\,\,\,\,\,\,\,  \,\,\,\,\,\,   \,\,\,\,  \,\,\,\,\,\,   (80 GeV protons, 1.07 MW power, 1.47$\times 10^{21}$ POT/yr, 40 kt fiducial mass, 3.5+3.5 yr)  & app: 2.0\% \,\,\,\, \,\,\,\,\,\,\,\quad\quad dis: 5.0\% & app: 5-20\%\,\, \,\,\,\,\,\,\,\,\,\quad\quad dis: 5-20\% \\
\hline
T2HK \quad(WC)& $\dfrac{295}{0.6}$ & 864.5 + 2593.5\,\,\,\,\,\,\,\,\,\,\,\,\,\,\,\,\,\,\,\,\,\,\,\,\,\,\,\,  \,\,\,\,\,\,   \,  \,\,\,\,\,\,   (30 GeV protons, 1.3 MW power, 2.7$\times 10^{21}$ POT/yr, 0.19 Mt each tank, 1.5+4.5 yr with 1 tank, 1+3 yr with 2 tanks)  & app: 2.5\% \,\,\, \,\,\,\,\,\,\,\,\quad dis: 2.5\% & app: 5\% \,\,\,\,\, \,  \,\,\,\,\,\, \,\,\,\,\,\, \,\,\,\,\,\,\,\quad\quad dis: 20\% \\
\hline
\vspace{9 mm}T2HKK-1.5\quad(WC)\vspace{9 mm}  & $\dfrac{295}{0.6}+ \dfrac{1100}{0.8}$ & \multirow{2}{5cm}{1235 + 3705 \quad\quad \quad\quad \quad(30 GeV protons, 1.3 MW power, 2.7$\times 10^{21}$ POT/yr, 0.19 Mt each tank, 2.5+7.5 yr with 1 tank at KD and HK)} & \multirow{2}{2.2cm}[-2pt]{app: 2.5\% \,\,\,\,\,\,\quad\quad dis: 2.5\%} &\multirow{2}{2.2cm}{app: 5\% \,\,\,\,\,\quad\quad dis: 20\% }\\
\cline{1-2}
\vspace{9 mm}T2HKK-2.5\quad(WC)\vspace{9 mm} & $\dfrac{295}{0.6}+ \dfrac{1100}{0.6}$ & & & \\
\hline
\end{tabular}
\end{center}
{\footnotesize For DUNE, 1 yr $= 1.76\times 10^7$s; for HyperK, 1 yr $= 1.0\times 10^7$s.}
\label{tb:exp}
\end{table}

\subsection{Simulation details}

\begin{figure}
\centering
\includegraphics[width=1.0\textwidth]{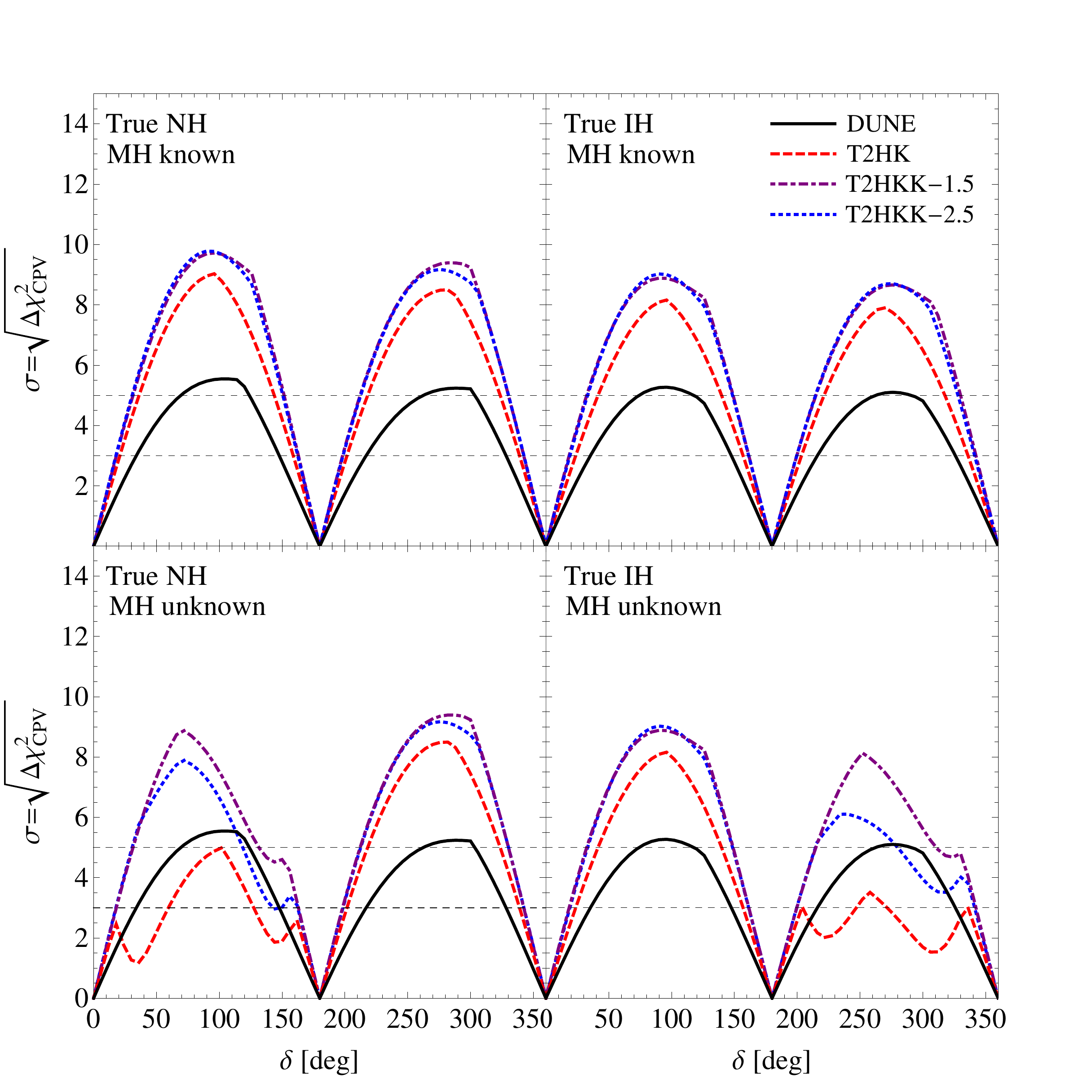}
\caption{The expected sensitivity to CP violation as a function of SM $\delta$ at DUNE, T2HK, and T2HKK. The central values and uncertainties of the oscillation parameters are adopted from a global fit in the SM scenario~\cite{Gonzalez-Garcia:2014bfa}, and a 5\% uncertainty for the matter density is assumed.}
\label{fig:cpv}
\end{figure}

\begin{figure}
\centering
\includegraphics[width=1.0\textwidth]{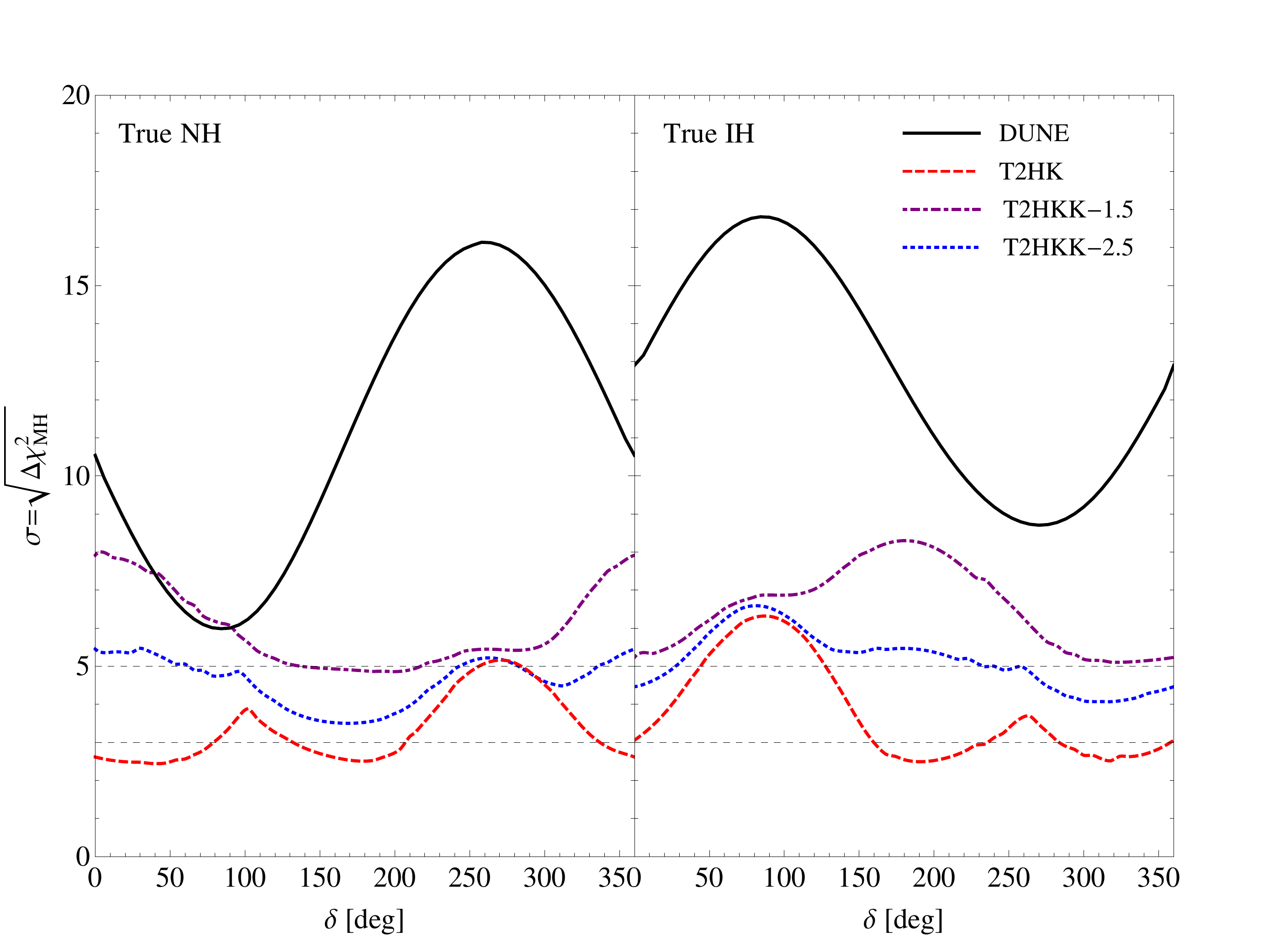}
\caption{The expected sensitivity to the mass hierarchy as a function
of SM $\delta$ at DUNE, T2HK, and T2HKK. The inputs and assumptions are the
same as in Fig.~\ref{fig:cpv}.
}
\label{fig:MH}
\end{figure}

We simulate the experiments using the GLoBES
software~\cite{GLOBES}. We use the official GLoBES simulation files
released by the DUNE collaboration~\cite{Alion:2016uaj} which has the
same experimental configurations as the DUNE CDR. The
normalization uncertainties for the appearance and disappearance signal
rates are 2\% and 5\%, respectively. The background uncertainties are
5\% except for the $\nu_\tau$ background, which has a 20\%
uncertainty. For the T2HK and T2HKK experiments, we matched the number
of events reported in Tables XXIX and XXX of Ref.~\cite{HK} in our
simulation. We assume a normalization uncertainty of 2.5\% for the
signal rates, and 5\% (20\%) for the appearance (disappearance)
background rates. Using the central values and uncertainties from a
global fit in the SM scenario~\cite{Gonzalez-Garcia:2014bfa}, we show
the expected CP violation and mass hierarchy sensitivity of DUNE,
T2HK, and T2HKK as a function of $\delta$ for both true normal and true
inverted hierarchies in Figs.~\ref{fig:cpv} and~\ref{fig:MH}, respectively.
From Fig.~\ref{fig:cpv}, we see that the expected CP violation sensitivities
in our simulation are consistent with those in the
DUNE~\cite{Acciarri:2015uup} and Hyper-K~\cite{HK} design reports.

For the NSI scenario, we use the new physics tools developed in
Refs.~\cite{Kopp:2007ne, Kopp:2008ds}. In our simulation, we use the
Preliminary Reference Earth Model density profile~\cite{Dziewonski:1981xy}
with a 5\% uncertainty for the matter density.\footnote{Note that in the
DUNE CDR~\cite{Alion:2016uaj} a 2\% uncertainty is
used for the matter density, while a 6\% uncertainty is used in the T2HK~\cite{HK}
and T2HKK~\cite{T2HKK} reports.}
The central values and uncertainties for the
mixing angles and mass-squared differences are adopted from the
global fit with NSI in Ref.~\cite{Gonzalez-Garcia:2013usa}, which are
\bea
&\sin^2\theta_{13}=0.023\pm0.002\,,\quad \sin^2\theta_{23}=0.43^{+0.08}_{-0.03}\,,\nonumber
\\
&\sin^2\theta_{12}=(0.305\pm0.015)\oplus(0.70\pm0.017)\,,\nonumber
\\
&\delta m_{21}^2=(7.48\pm0.21)\times 10^{-5} \text{eV}^2\,, \quad |\delta m_{31}^2|=(2.43\pm0.08)\times 10^{-3} \text{eV}^2\,.
\eea

For the NSI parameters, we scan over the following ranges suggested by the
analysis of Ref.~\cite{Gonzalez-Garcia:2013usa},
\bea
&-5.0<\epsilon_{ee}<5.0\,,\quad \epsilon_{e\mu}<0.5\,,\quad \epsilon_{e\tau}<1.2\,,
\\\nonumber
& -0.6<\epsilon_{\tau\tau}<0.6\,,\quad \epsilon_{\mu\tau}<0.1\,,
\label{eq:eps-range}
\eea
and marginalize over all the NSI phases in our simulation.

\section{Sensitivities to NSI parameters and CP violation}

\subsection{Oscillation probabilities}

 The appearance probability for the normal hierarchy (NH) can be written
as~\cite{Liao:2016hsa}
\bea
P(\nu_\mu \to \nu_e) &= &x^2 f^2 + 2xyfg \cos(\Delta + \delta) + y^2 g^2
\nonumber\\
&+& 4\hat A \epsilon_{e\mu}
\left\{ xf [s_{23}^2 f \cos(\phi_{e\mu}+\delta)  
+ c_{23}^2 g \cos(\Delta+\delta+\phi_{e\mu})]\right.
\nonumber\\
&\phantom{+}&\qquad\qquad \left.
+yg [c_{23}^2 g \cos\phi_{e\mu} + s_{23}^2 f \cos(\Delta-\phi_{e\mu})]\right\}
\nonumber\\
&+& 4\hat A \epsilon_{e\tau} s_{23} c_{23}
\left\{ xf [f \cos(\phi_{e\tau}+\delta)  
- g \cos(\Delta+\delta+\phi_{e\tau})] \right.
\nonumber\\
&\phantom{+}&\qquad\qquad\qquad \left.
-yg [g \cos\phi_{e\tau} - f \cos(\Delta-\phi_{e\tau})]\right\}
\nonumber\\
&+& 4 \hat A^2 \left( g^2 c_{23}^2 |c_{23} \epsilon_{e\mu} - s_{23}\epsilon_{e\tau}|^2
 +   f^2 s_{23}^2 |s_{23} \epsilon_{e\mu} + c_{23}\epsilon_{e\tau}|^2 \right)
\nonumber\\
&+& 8 \hat A^2 fg s_{23} c_{23}
\left\{ c_{23}\cos\Delta
\left[ s_{23}(\epsilon_{e\mu}^2 - \epsilon_{e\tau}^2)+2 c_{23} \epsilon_{e\mu}\epsilon_{e\tau}
\cos(\phi_{e\mu}-\phi_{e\tau})\right]\right.
\nonumber\\
&\phantom{+}& \qquad\qquad\qquad  \left.-\epsilon_{e\mu}\epsilon_{e\tau}
\cos(\Delta-\phi_{e\mu}+\phi_{e\tau})\right\}
+ {\cal O}(s_{13}^2 \epsilon, s_{13}\epsilon^2, \epsilon^3)\,,
\label{eq:app-prob}
\eea
where following Ref.~\cite{Barger:2001yr},
\bea
x &\equiv& 2 s_{13} s_{23}\,,\quad
y \equiv 2r s_{12} c_{12} c_{23}\,,
\quad r = |\delta m^2_{21}/\delta m^2_{31}|\,,
\nonumber\\
f,\, \bar{f} &\equiv& \frac{\sin[\Delta(1\mp\hat A(1+\epsilon_{ee}))]}{(1\mp\hat A(1+\epsilon_{ee}))}\,,\ 
g \equiv \frac{\sin(\hat A(1+\epsilon_{ee}) \Delta)}{\hat A(1+\epsilon_{ee})}\,,\nonumber\\
\Delta &\equiv &\left|\frac{\delta m^2_{31} L}{4E}\right|,\ 
\hat A \equiv \left|\frac{A}{\delta m^2_{31}}\right|\,.
\label{eq:define}
\eea
Henceforth we define $P_{\mu e}\equiv P(\nu_\mu \to \nu_e)$ and the
antineutrino probability $\overline{P}_{\mu e}\equiv
P(\overline{\nu}_\mu \to \overline{\nu}_e)$, which is given by
Eq.~(\ref{eq:app-prob}) with $\hat A \to - \hat A$ (and hence $f \to
\bar{f}$), $\delta \to - \delta$, and $\phi_{\alpha\beta} \to -
\phi_{\alpha\beta}$. For the inverted hierarchy (IH), $\Delta \to -
\Delta$, $y \to -y$, $\hat A \to - \hat A$ (i.e., $f \leftrightarrow -
\bar{f}$, and $g \to -g$). Our result agrees with the ${\cal
  O}(\epsilon)$ expressions in Refs.~\cite{Kopp:2007ne, Kikuchi:2008vq}.

From Eq.~(\ref{eq:app-prob}), we see that $\epsilon_{\mu\mu}$,
$\epsilon_{\mu\tau}$ and $\epsilon_{\tau\tau}$ do not appear in the
appearance probability up to second order in $\epsilon$. Hence, they
mainly affect the disappearance channel. Taking $\epsilon_{ee}$,
$\epsilon_{e\mu}$ and $\epsilon_{e\tau}$ equal to zero, the disappearance
probability can be written as
\bea
P(\nu_\mu \to \nu_\mu) &=& 1- \sin^22\theta_{23}\sin^2\Delta+4r c_{12}^2c_{23}^2s_{23}^2\Delta\sin2\Delta-\frac{4s_{23}^4s_{13}^2\sin^2(1-\hat{A})\Delta}{(1-\hat{A})^2}
\nonumber\\
&-&\frac{\sin^22\theta_{23}s_{13}^2}{(1-\hat{A})^2}\left[\hat{A}(1-\hat{A})\Delta\sin2\Delta+\sin(1-\hat{A})\Delta\sin(1+\hat{A})\Delta\right]
\nonumber\\
&-& 2 \hat{A}\epsilon_{\mu\tau}\cos\phi_{\mu\tau}(\sin^32\theta_{23}\Delta\sin2\Delta+2\sin2\theta_{23}\cos^2 2\theta_{23}\sin^2\Delta)
\nonumber\\
&+&  \hat{A}(\epsilon_{\mu\mu}-\epsilon_{\tau\tau})\sin^22\theta_{23}\cos2\theta_{23}\left(\Delta\sin2\Delta-2\sin^2\Delta \right)
\nonumber\\
&-&2\hat{A}^2\sin^2 2\theta_{23} \epsilon_{\mu\tau}^2 \left(2\sin^2 2\theta_{23}\cos^2\phi_{\mu\tau}\Delta^2\cos2\Delta
+\sin^2\phi_{\mu\tau}\Delta  \sin2\Delta\right)
\nonumber\\
&-&\hat{A}^2\sin^4 2\theta_{23} (\epsilon_{\mu\mu}-\epsilon_{\tau\tau})^2(\frac{1}{2}\Delta\sin2\Delta-\sin^2\Delta)
\nonumber\\
&+& {\cal O}(s_{13}^2\epsilon, r\epsilon, s_{13}\epsilon^2, \cos2\theta_{23}\epsilon^2, \epsilon^3)\,.
\label{eq:dis-prob}
\eea
Our result agrees with Ref.~\cite{Asano:2011nj} for the SM terms (in the
first two lines) and with Ref.~\cite{Kikuchi:2008vq} for the NSI terms up
to second order after making the assumption that terms of order
$\cos2\theta_{23} \epsilon^2$
can be ignored.  Our result disagrees with Ref.~\cite{Coloma:2015kiu}
in the second-order terms in $\epsilon$.  We can see in
Eq.~(\ref{eq:dis-prob}) that $\epsilon_{\mu\mu}$ and
$\epsilon_{\tau\tau}$ appear in the form of their difference up to
second order in $\epsilon$. We therefore choose $\epsilon_{\mu\mu}=0$.

\subsection{NSI in the appearance channels ($\epsilon_{e\mu}$, $\epsilon_{e\tau}$ and $\epsilon_{ee}$)}

We only consider $\epsilon_{e\mu}$, $\epsilon_{e\tau}$ and
$\epsilon_{ee}$ in this section because $\epsilon_{\mu\mu}$, $\epsilon_{\mu\tau}$ and
$\epsilon_{\tau\tau}$ do not appear in the appearance probabilities up
to second order in $\epsilon$.

\subsubsection{A single nonzero NSI parameter}

For a single $L/E$, data consistent with the SM can be also
described by a model with NSI if
$P^\text{SM}(\nu_\mu\rightarrow\nu_e)=P^\text{NSI}(\nu_\mu\rightarrow\nu_e)$
and
$P^\text{SM}(\bar{\nu}_\mu\rightarrow\bar{\nu}_e)=P^\text{NSI}(\bar{\nu}_\mu\rightarrow\bar{\nu}_e)$.
Since the three mixing angles, $\delta m_{21}^2$ and $|\delta m_{31}^2|$ are
well-measured by other experiments, if only one off-diagonal NSI
($\epsilon_{e\mu}$ or $\epsilon_{e\tau}$) is nonzero, there exists a
continuous four-fold degeneracy as a result of the unknown mass hierarchy and
$\theta_{23}$ octant~\cite{Liao:2016hsa}.

The continuous degeneracy can be understood as follows. If only one off-diagonal NSI is nonzero, there are three unknowns to be
determined in the NSI scenario: $\delta^\prime$
(the Dirac CP phase in $P^\text{NSI}$), the NSI magnitude $\epsilon$
and the NSI phase $\phi$. Since a single measurement of $P$ and
$\overline{P}$ for a fixed $L$ and $E$ gives only two constraints, for each value of
$\delta$ in the SM, a solution for $\epsilon$ and $\phi$ will exist for {\it
any} value of $\delta^\prime$. 
This leads to continuous degeneracies throughout the two-dimensional $\delta$-$\delta^\prime$ space.  
An additional measurement at a different
$L$ and/or $E$ can be made to reduce the degeneracies to lines in
$\delta$-$\delta^\prime$ space, i.e., for each value of $\delta$ there
will only be one $\delta^\prime$ that will be degenerate. If
there are multiple $\delta^\prime$ solutions, then a second additional measurement at a different $L$ and/or $E$ should in principle remove the degeneracies. 

If only $\epsilon_{ee}$ is nonzero, since it is real, an experiment that measures $P$ and $\overline{P}$
at a single $L/E$ should be able to fix the SM value of $\delta$ and the NSI values of
$\delta^\prime$ and $\epsilon_{ee}$. If a nontrivial solution exists, then there is a simple two-fold degeneracy between the SM and NSI, and at least one additional measurement
is needed to break the degeneracy between the SM and NSI with $\epsilon_{ee}$. Note however that the nonlinearity (in $\epsilon_{ee}$) of the equations may yield several solutions with nonzero $\epsilon_{ee}$ and $\delta^\prime\ne\delta$.

\begin{figure}
\centering
\includegraphics[width=1.0\textwidth]{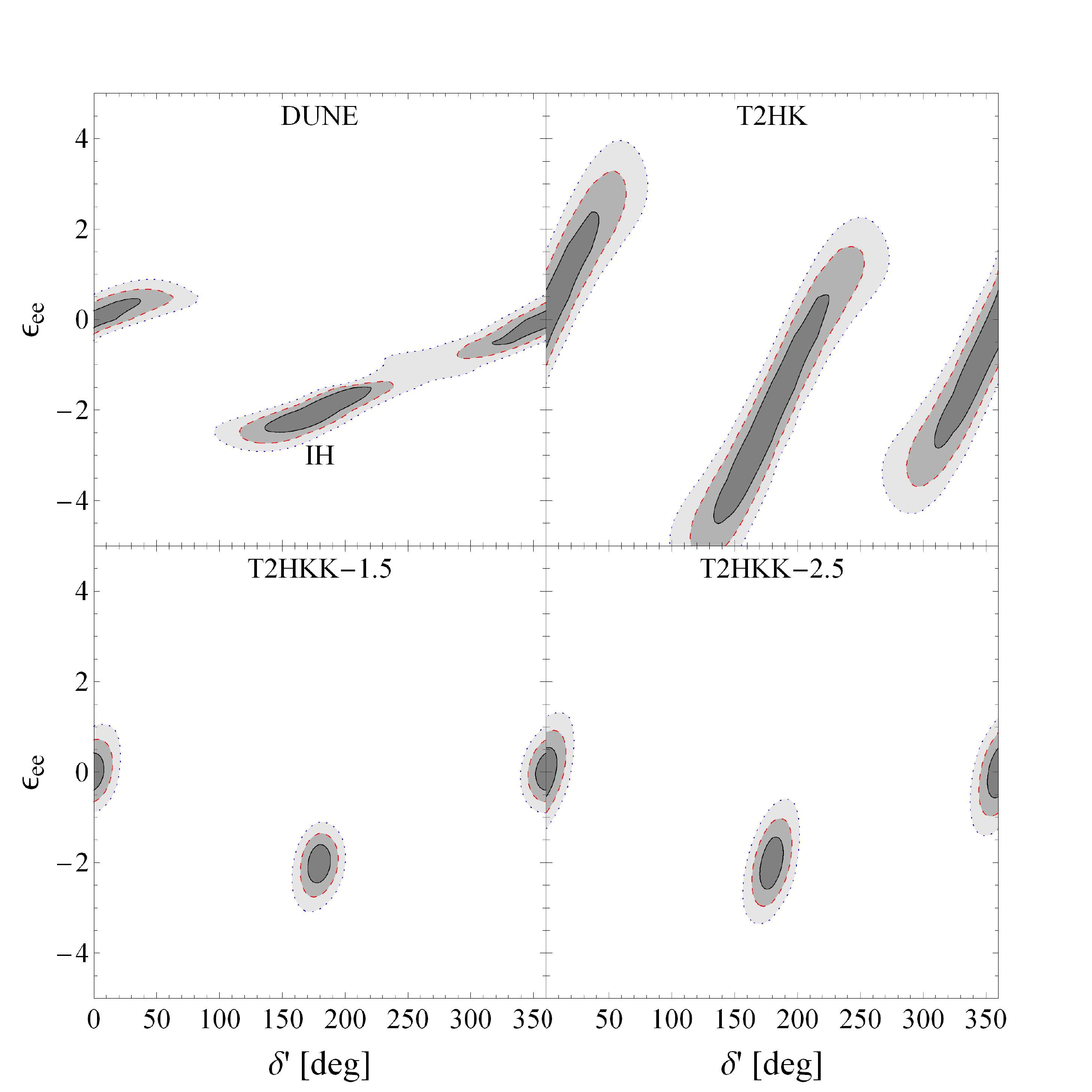}
\caption{$1\sigma$, $2\sigma$ and $3\sigma$ allowed regions for 
$\epsilon_{ee}$ at DUNE, T2HK, and T2HKK when only $\epsilon_{ee}$ is
nonzero. The data are consistent with the SM with $\delta = 0$ and the
NH. The allowed regions near $\delta^\prime=180^\circ$ are for the IH.
}
\label{fig:ee-1eps}
\end{figure}

\begin{figure}
\centering
\includegraphics[width=1.0\textwidth]{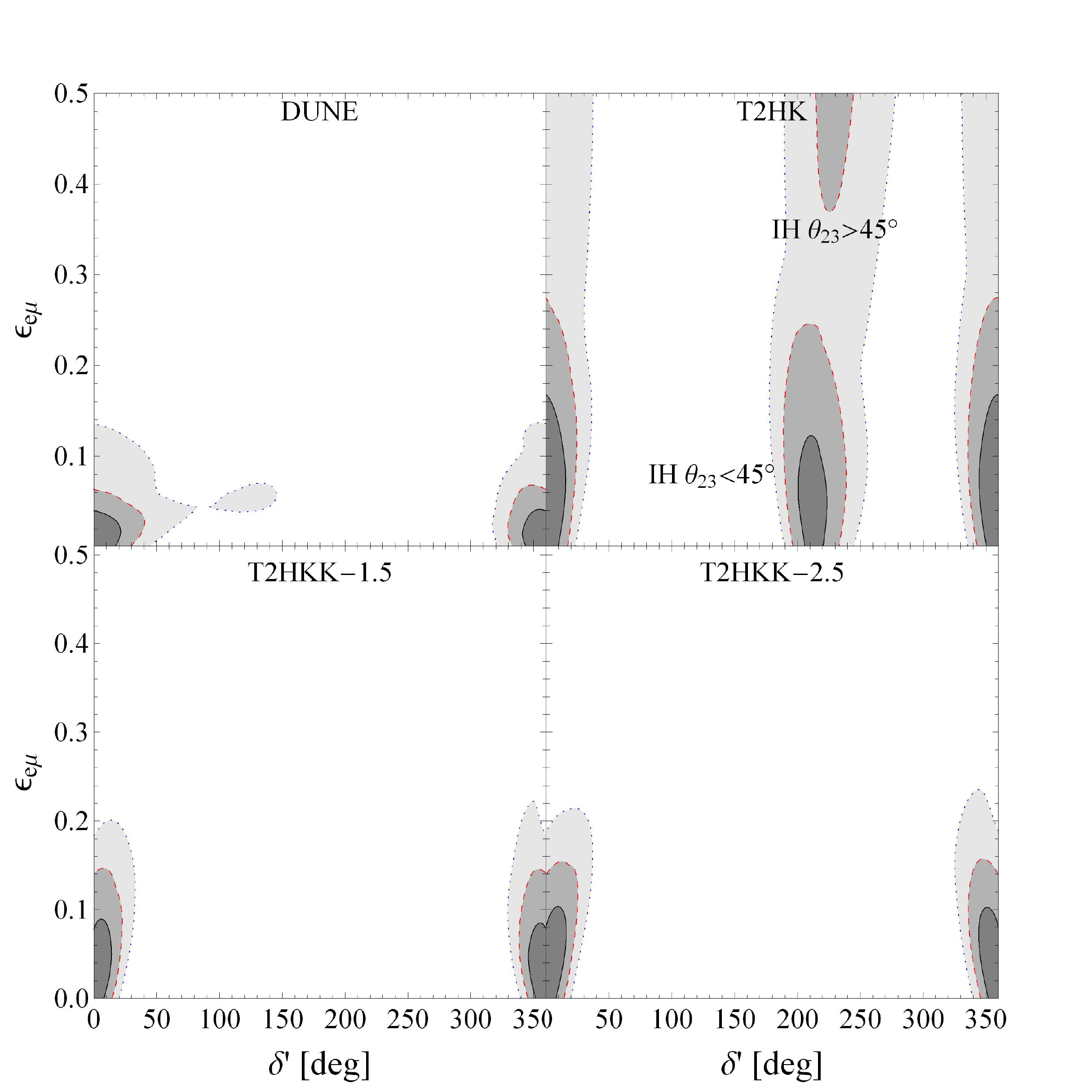}
\caption{Same as Fig.~\ref{fig:ee-1eps}, except for $\epsilon_{e\mu}$.
}
\label{fig:em-1eps}
\end{figure}

\begin{figure}
\centering
\includegraphics[width=1.0\textwidth]{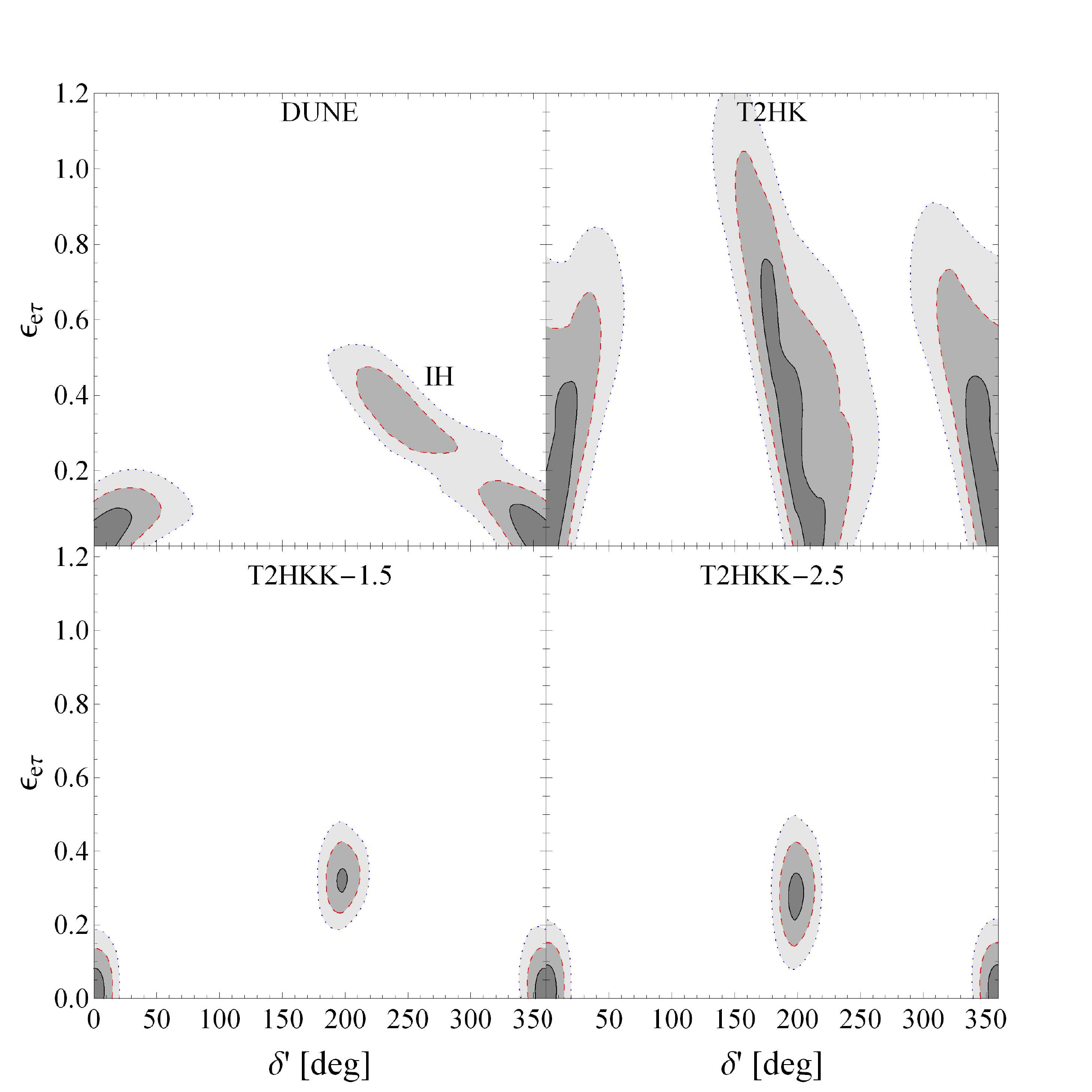}
\caption{Same as Fig.~\ref{fig:ee-1eps}, except for $\epsilon_{e\tau}$.
}
\label{fig:et-1eps}
\end{figure}

Since DUNE, T2HK and T2HKK effectively measure probabilities at a
variety of energies, in principle these experiments can not only
resolve the degeneracies with NSI solutions, but also put severe
restrictions on the NSI parameters.  If only one NSI parameter is nonzero, the
expected allowed regions in the 
$\delta^\prime-\epsilon_{ee}$, $-\epsilon_{e\mu}$ and $-\epsilon_{e\tau}$ planes are
shown in Figs.~\ref{fig:ee-1eps}, \ref{fig:em-1eps} and
\ref{fig:et-1eps}, respectively. We assume the data are consistent with
the SM with $\delta = 0$ and the NH. The results
are obtained after scanning over both mass hierarchies.

From Fig.~\ref{fig:ee-1eps}, we see that there is always an allowed
region near $\epsilon_{ee} = -2$ and $\delta'=180^\circ$. This
degeneracy at DUNE was first shown in Ref.~\cite{Liao:2016hsa}, and
can be explained by the generalized MH
degeneracy~\cite{Coloma:2016gei, Bakhti:2014pva}, which states that
under the transformation,
\bea
&\delta m_{31}^2\rightarrow -\delta m_{32}^2,\quad \theta_{12}\rightarrow 90^\circ-\theta_{12},\quad\delta\rightarrow 180^\circ-\delta,
\\\nonumber
&\epsilon_{ee}\rightarrow-\epsilon_{ee}-2,\quad \epsilon_{\alpha\beta}e^{i \phi_{\alpha\beta}}\rightarrow-\epsilon_{\alpha\beta}e^{-i \phi_{\alpha\beta}}\,\,
(\alpha\beta\neq ee)\,,
\eea
the Hamiltonian transforms as $H\rightarrow-H^*$, and the
oscillation probabilities are unchanged~\cite{Coloma:2016gei}. Since
this degeneracy does not depend on $L$ and $E$, all
long-baseline experiments, including atmospheric and reactor neutrino experiments (like JUNO~\cite{An:2015jdp})
can not resolve this degeneracy if $\epsilon_{ee}\sim -2$.

From Fig.~\ref{fig:em-1eps}, we see that if only $\epsilon_{e\mu}$ is
nonzero, the mass hierarchy degeneracy is resolved at DUNE and
T2HKK. DUNE puts severe constraints on $\epsilon$ ($ \lsim 0.15$ at
$3\sigma$) while T2HKK-1.5 places better constraints on $|\delta^\prime|$
($\lsim 30^\circ$ at $3\sigma$).  However, T2HK cannot resolve the mass
hierarchy in this case; the IH is still allowed for $\delta^\prime
\sim 215^\circ$. Also, there is a 2$\sigma$ allowed region around
$\epsilon_{e\mu}\sim 0.5$ arising from the $\theta_{23}$ octant
degeneracy. Around this region, the second octant of $\theta_{23}$ for
the IH has a smaller $\chi^2$ than the first octant.

If only $\epsilon_{e\tau}$ is nonzero, from Fig.~\ref{fig:et-1eps}
we see that the mass hierarchy is not resolved for any of the
experiments, although the IH is not allowed at 
the $1\sigma$ CL at DUNE. This could lead to a wrong determination of
the Dirac CP phase in all the experiments.


\subsubsection{Three nonzero NSI parameters}

If $\epsilon_{e\mu}$, $\epsilon_{e\tau}$ and $\epsilon_{ee}$ are all
nonzero, then there are six free NSI parameters: $\delta^\prime$,
$\epsilon_{ee}$, two magnitudes, and two phases. Even $P$ and
$\overline{P}$ measurements at three different $L$ and $E$
combinations (six equations and six unknowns) could at most reduce the
degeneracy to a single point in NSI parameter space (or perhaps a
finite number of points). Therefore, an experiment that measures
probabilities at a large variety of energies and/or distances is
needed to resolve the degeneracies in the presence of multiple NSI.

\begin{figure}
\centering
\includegraphics[width=1.0\textwidth]{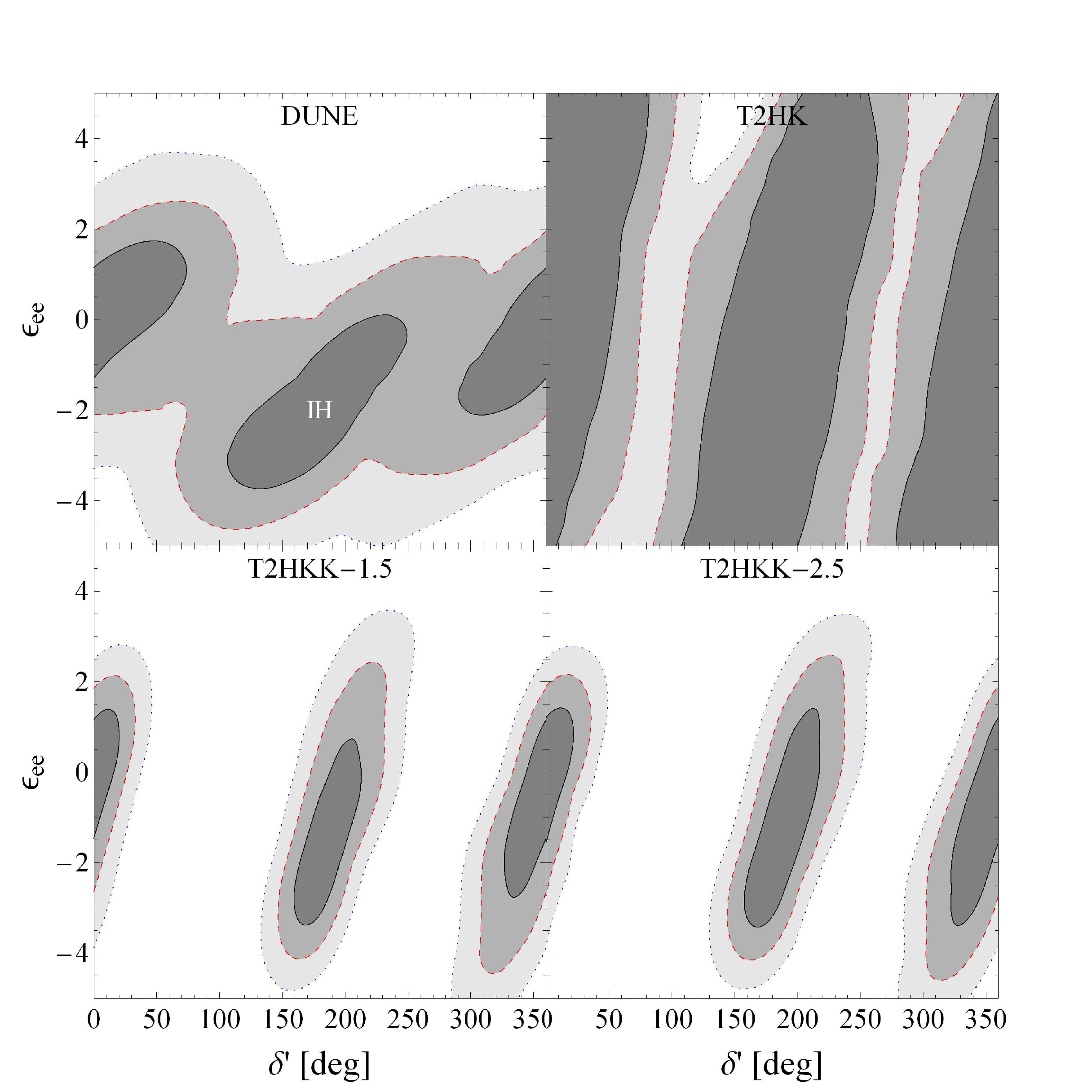}
\caption{$1\sigma$, $2\sigma$ and $3\sigma$ allowed regions for
$\epsilon_{ee}$ at DUNE, T2HK and T2HKK when $\epsilon_{e\mu}$,
$\epsilon_{e\tau}$ and $\epsilon_{ee}$ are all nonzero. The data are
consistent with the SM with $\delta = 0$ and the NH. All other
parameters not shown have been marginalized over.  }
\label{fig:ee-3eps}
\end{figure}

\begin{figure}
\centering
\includegraphics[width=1.0\textwidth]{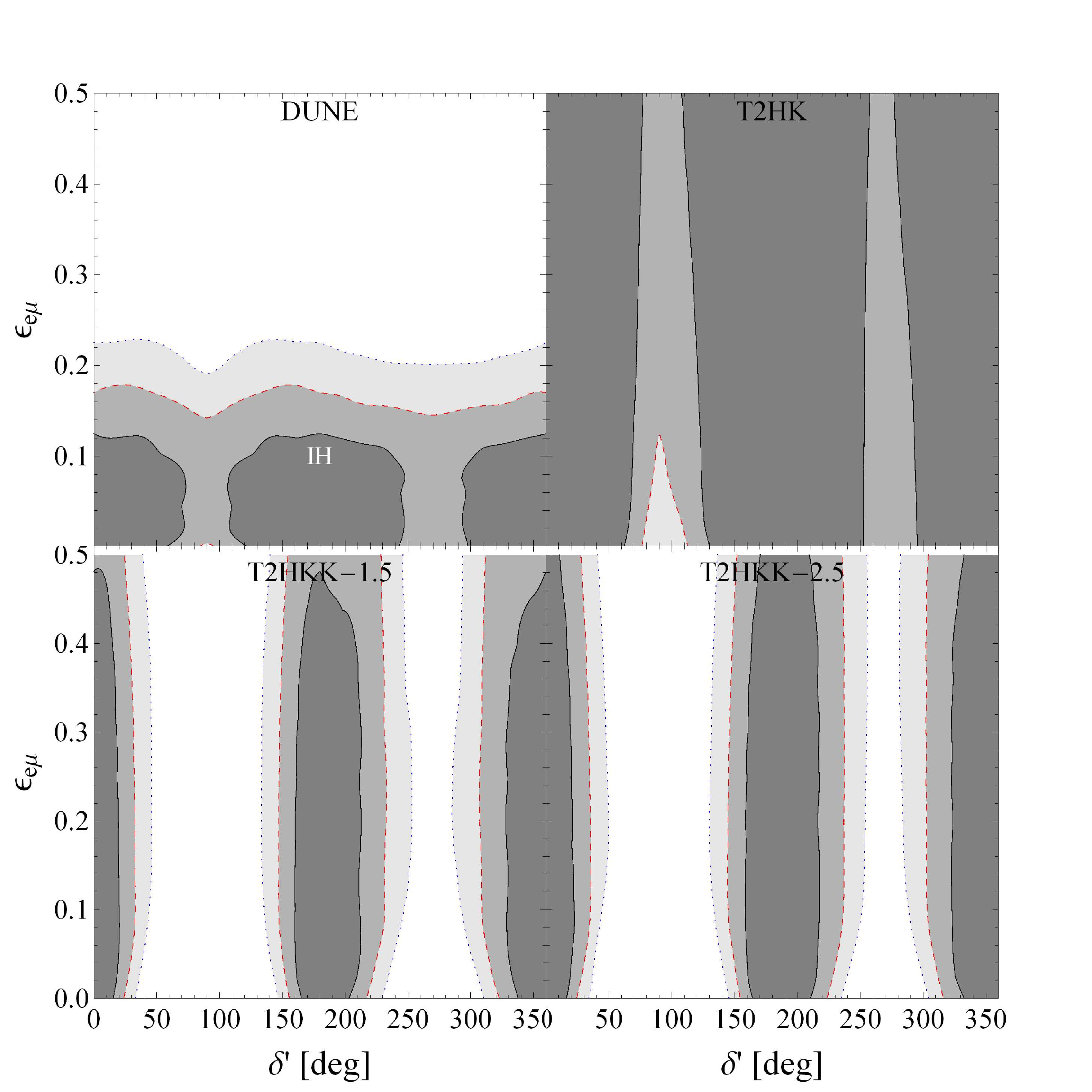}
\caption{Same as Fig.~\ref{fig:ee-3eps}, except for $\epsilon_{e\mu}$.
}
\label{fig:em-3eps}
\end{figure}

\begin{figure}
\centering
\includegraphics[width=1.0\textwidth]{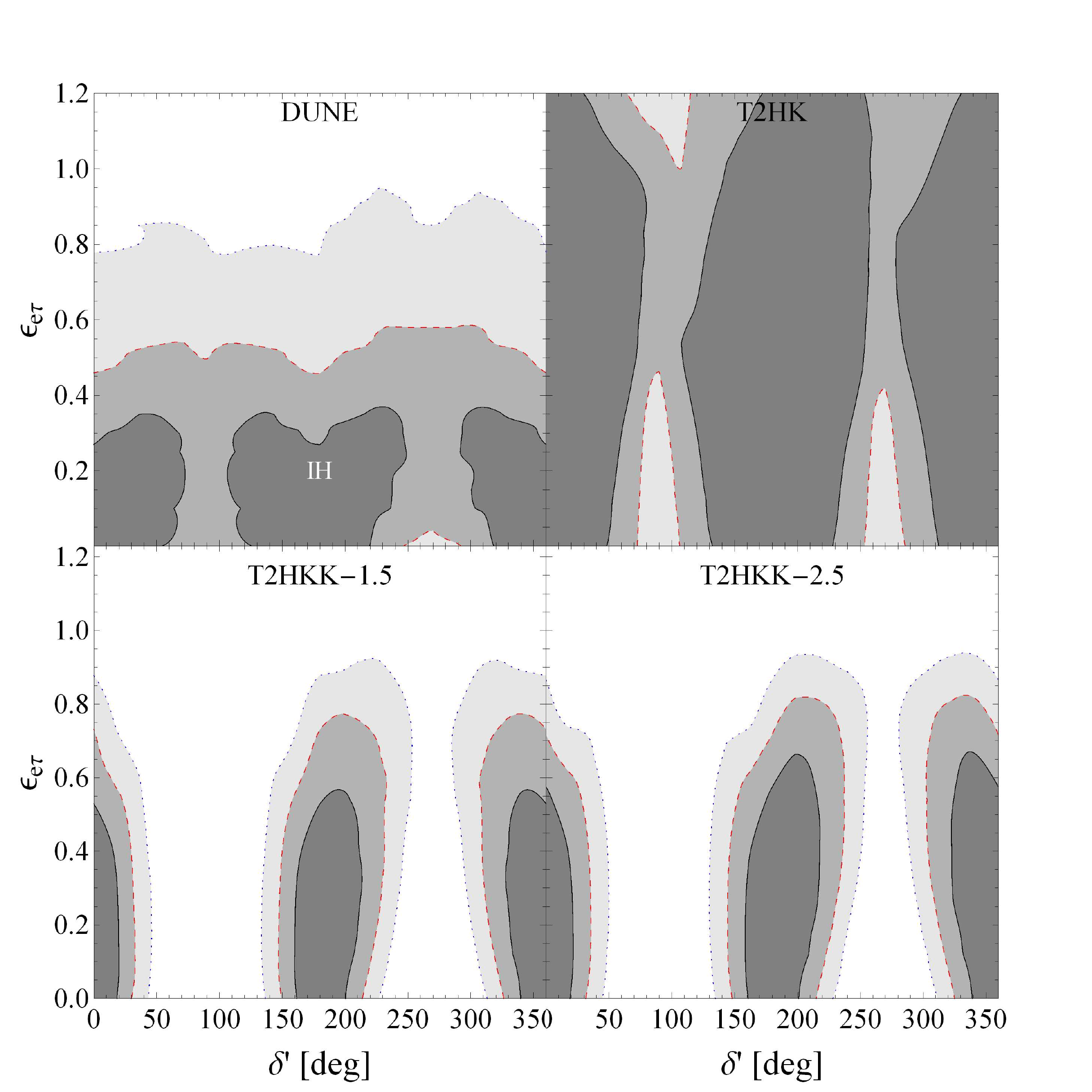}
\caption{Same as Fig.~\ref{fig:ee-3eps}, except for $\epsilon_{e\tau}$.
}
\label{fig:et-3eps}
\end{figure}

\begin{figure}
\centering
\includegraphics[width=1.0\textwidth]{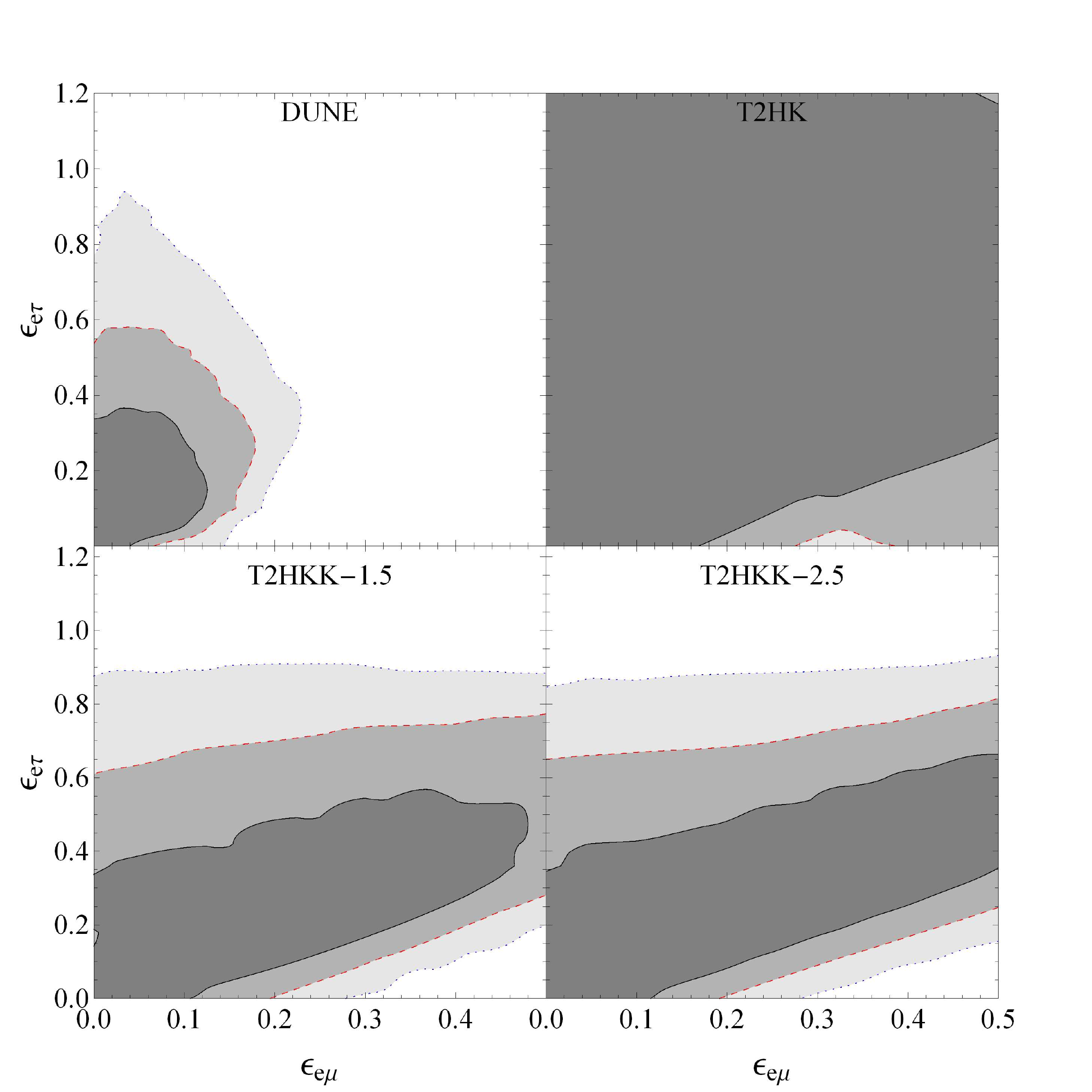}
\caption{$1\sigma$, $2\sigma$ and $3\sigma$ allowed regions in the
  $\epsilon_{e\mu}-\epsilon_{e\tau}$ plane at DUNE, T2HK and
  T2HKK, assuming $\epsilon_{ee}$, $\epsilon_{e\mu}$ and $\epsilon_{e\tau}$
  are all nonzero. The data are consistent with the SM and the NH with $\delta =
  0$, and all parameters not shown have been marginalized over.  }
\label{fig:em-et-3eps}
\end{figure}

The expected allowed regions in the
$\delta^\prime-\epsilon_{ee}$, $-\epsilon_{e\mu}$ and $-\epsilon_{e\tau}$ planes are
shown in Figs.~\ref{fig:ee-3eps},~\ref{fig:em-3eps}
and~\ref{fig:et-3eps}, respectively. For each $\epsilon$, we scan over
both NH and IH, and marginalize over all the other NSI parameters. As expected,
 constraints on the NSI parameters become much worse. 
 In particular, from Fig.~\ref{fig:em-3eps},
we see that the constraint on $\epsilon_{e\mu}$ is much weaker at T2HK
and T2HKK than at DUNE. This coincides with a strong degeneracy
between $\epsilon_{e\mu}$ and $\epsilon_{e\tau}$ at T2HK and T2HKK
(see Fig.~\ref{fig:em-et-3eps}), and can be explained by examining the
appearance probability in Eq.~(\ref{eq:app-prob}).

For T2HK and T2HKK, since $\hat A \sim 0.05$ for $E \sim 0.6$~GeV, the
higher order terms from the matter effect can be neglected in
Eq.~(\ref{eq:app-prob}). Taking $\epsilon_{ee}=0$ and $\delta^\prime=\delta$,
we have 
\bea
P^\text{NSI}_{\mu e}-P^\text{SM}_{\mu e} &=& 4\hat A \epsilon_{e\mu}xf \left[s_{23}^2 f \cos(\phi_{e\mu}+\delta)  + c_{23}^2 g \cos(\Delta+\delta+\phi_{e\mu})\right]
\nonumber\\
&+& 4\hat A \epsilon_{e\tau}  xf \left[s_{23} c_{23}f \cos(\phi_{e\tau}+\delta)  - s_{23} c_{23} g \cos(\Delta+\delta+\phi_{e\tau})\right]
\nonumber\\
&+& {\cal O}(y \hat A \epsilon, \hat A^2\epsilon^2)\,.
\eea
If in addition, $\phi_{e\mu}=\phi_{e\tau}=\pm 90^\circ-\delta$, then
\bea
P^\text{NSI}_{\mu e}-P^\text{SM}_{\mu e} &=&
\mp\ 4xf\hat A c_{23} \left( c_{23} \epsilon_{e\mu}
- s_{23}\epsilon_{e\tau}\right) g \sin \Delta
+ {\cal O}(y \hat A \epsilon, \hat A^2\epsilon^2)\,.
\label{eq:Pdiff}
\eea
The corresponding equation for antineutrinos is given by
$\hat{A}\rightarrow-\hat{A}$, $\delta\rightarrow-\delta$ and
$\phi_{\alpha\beta}\rightarrow-\phi_{\alpha\beta}$. A similar
equation holds for the IH. As can be seen from Eq.~(\ref{eq:Pdiff}),
if
\be
\epsilon_{e\tau}=\cot\theta_{23}\epsilon_{e\mu}\,,
\label{eq:cancel}
\ee
the difference between the NSI and SM appearance probabilities is
strongly suppressed in both the neutrino and antineutrino
modes. Consequently, the constraint on $\epsilon_{e\mu}$ is
 very weak at T2HK and T2HKK if
 $\epsilon_{e\tau}\simeq \cot\theta_{23}\epsilon_{e\mu}$. 
 Since neutino energies at DUNE are much higher than at T2HK and
T2HKK (e.g., $E_\text{peak} \sim 3 $~GeV for which $\hat A \sim 0.28$),
higher order terms in Eq.~(\ref{eq:app-prob}) cannot be neglected,
and the degeneracy between $\epsilon_{e\mu}$ and $\epsilon_{e\tau}$
can be resolved. Also, comparing the lower panels of Fig.~\ref{fig:em-et-3eps} 
we see that T2HKK-1.5
starts to break the degeneracy between $\epsilon_{e\mu}$ and
$\epsilon_{e\tau}$ for $\epsilon_{e\mu}\lsim0.5$ since T2HKK-1.5 has a
higher peak energy than T2HKK-2.5.

\begin{figure}
\centering
\includegraphics[width=1.0\textwidth]{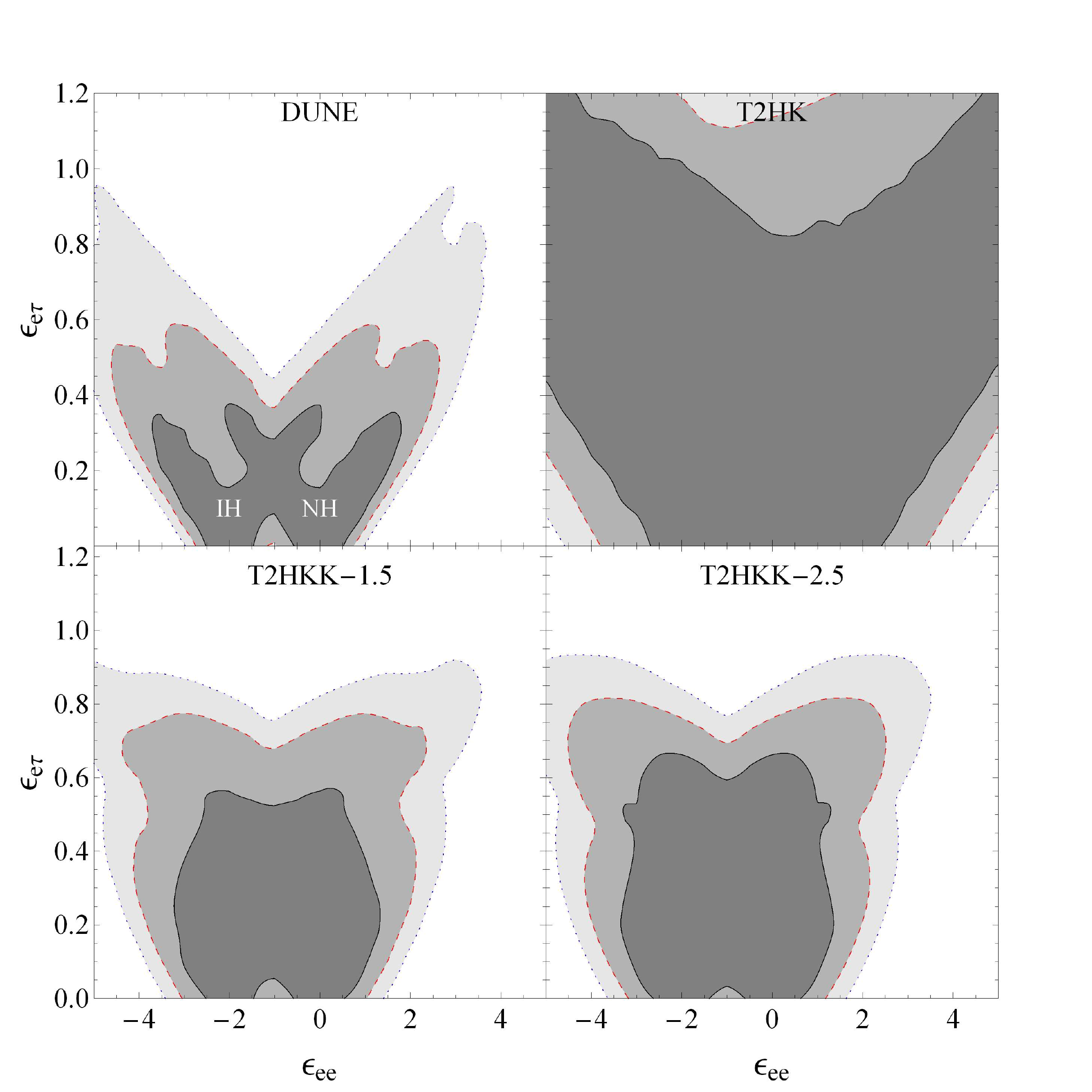}
\caption{Same as Fig.~\ref{fig:em-et-3eps}, except for $\epsilon_{e\tau}$ versus $\epsilon_{ee}$.
}
\label{fig:ee-et-3eps}
\end{figure}

We also find strong correlations between $\epsilon_{e\tau}$ and
$\epsilon_{ee}$ in all experiments, which can be seen in
Fig.~\ref{fig:ee-et-3eps}. The allowed regions are symmetric around
$\epsilon_{ee}=-1$ due to the generalized MH degeneracy; the vertex of the 
V-shaped NH region is at $\epsilon_{ee}=0$ and vertex of the V-shaped
 IH region is at $\epsilon_{ee}=-2$.


\subsubsection{Dependence of the sensitivity on $\delta$}

Since both the Dirac CP phase $\delta$ and the mass hierarchy are
unknown, the experimental performance may be affected by the true
parameters in nature. In this section, we examine how the sensitivity
changes with the true value of $\delta$. In the next section we
study the sensitivity if the true hierarchy is inverted.  Although
 the mass hierarchy will not be measured in neutrino
oscillation experiments because of the generalized MH degeneracy, future
neutrinoless double beta decay experiments may determine the mass
hierarchy if neutrinos are Majorana particles. We therefore entertain the
possibilities that the MH is known and that it is unknown.

\begin{figure}
\centering
\includegraphics[width=1.0\textwidth]{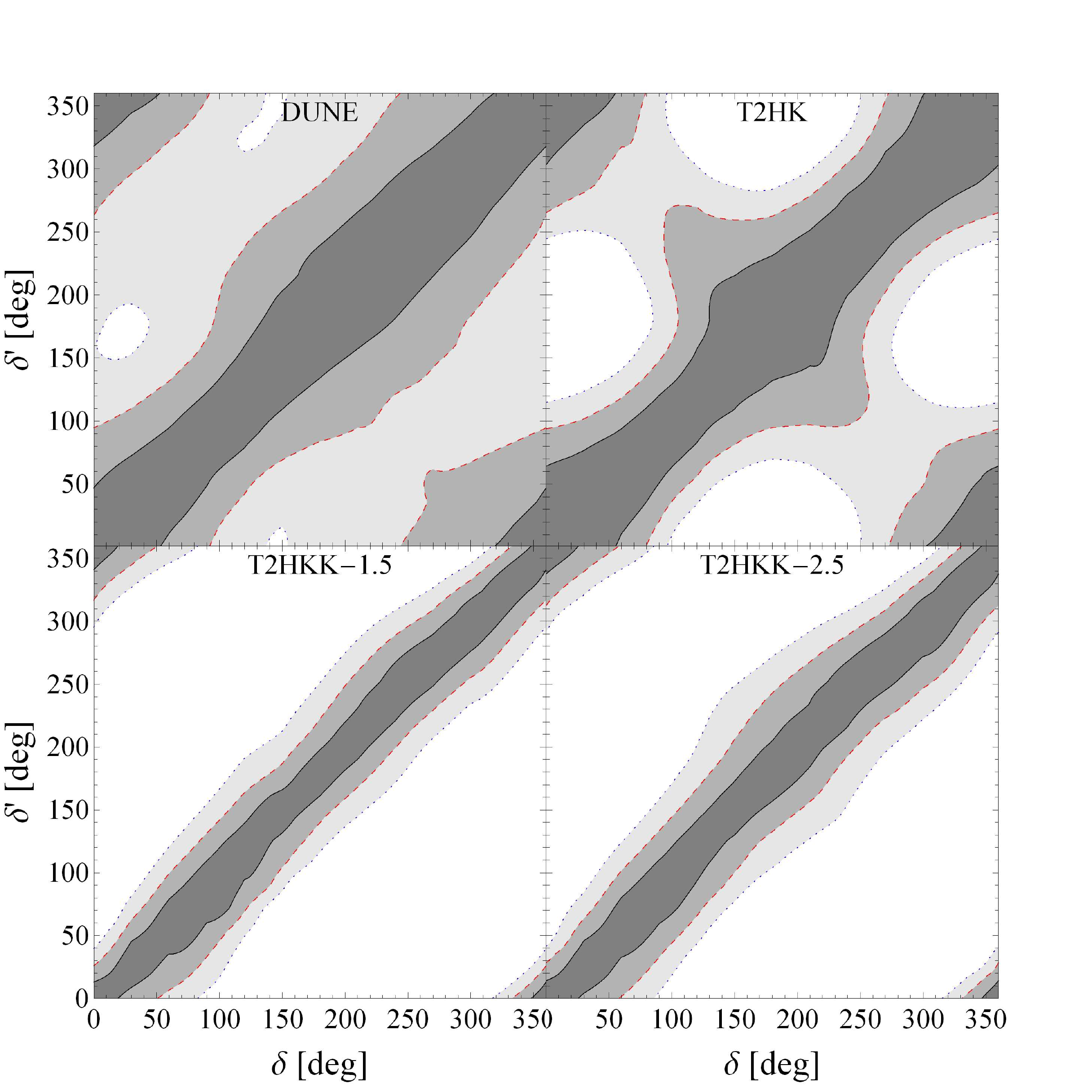}
\caption{$1\sigma$, $2\sigma$ and $3\sigma$ allowed regions for $\delta^\prime$ as a function of $\delta$ at DUNE, T2HK and T2HKK. The data are consistent with the SM and the NH. We assume the mass hierarchy is known, and $\epsilon_{ee}$, $\epsilon_{e\mu}$ and $\epsilon_{e\tau}$ are all nonzero. All parameters not shown have been marginalized over.
}
\label{fig:ddp-known}
\end{figure}

\begin{figure}
\centering
\includegraphics[width=1.0\textwidth]{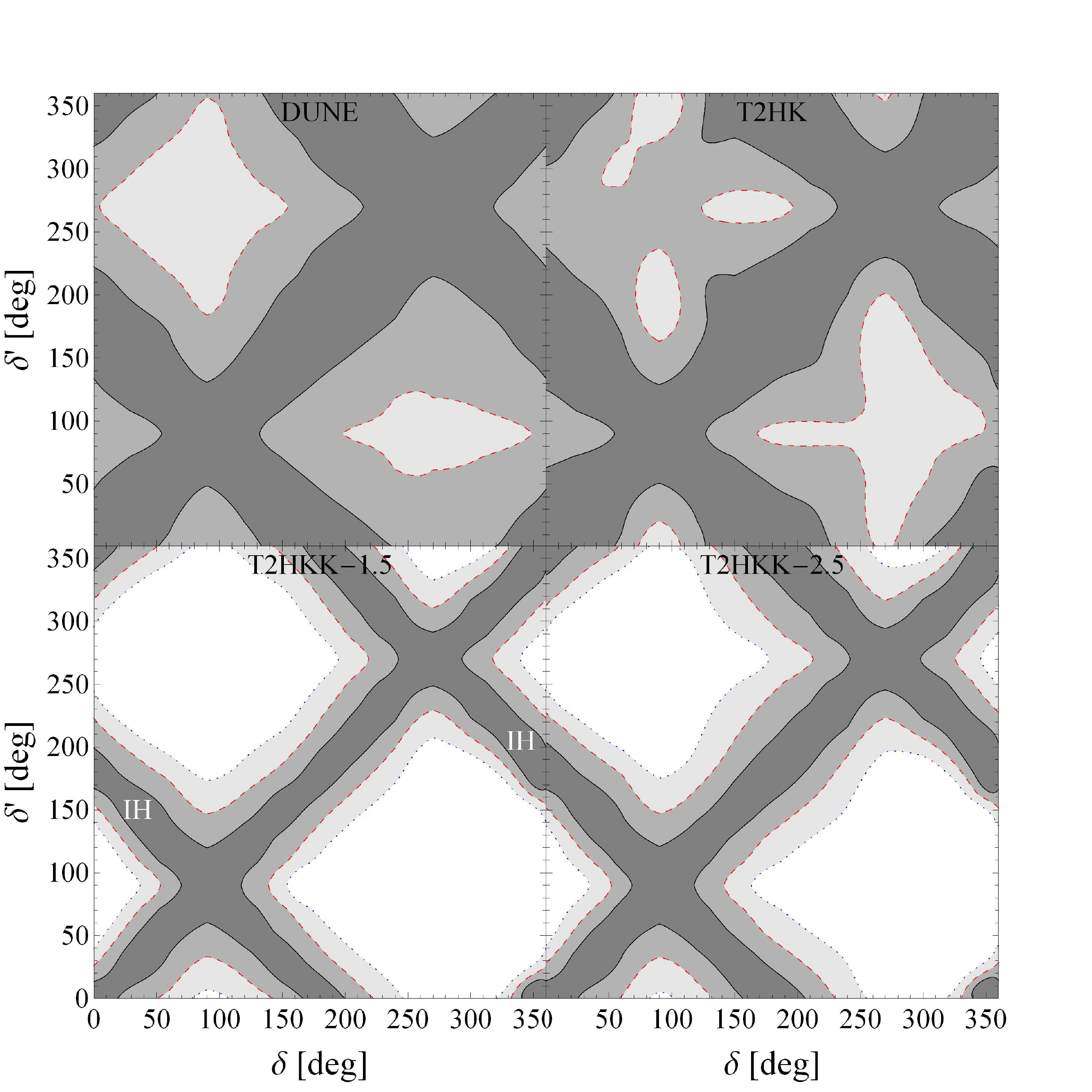}
\caption{Same as Fig.~\ref{fig:ddp-known}, except that the mass hierarchy is unknown. Note that the entire $\delta$-$\delta^\prime$ parameter space is allowed at 3$\sigma$ for T2HK and DUNE.
}
\label{fig:ddp-unknown}
\end{figure}

We assume that the data are consistent with the SM and the NH, and plot
the constraints on $\delta^\prime$ as a function of $\delta$ if all
three $\epsilon$'s are nonzero; see Figs.~\ref{fig:ddp-known}
and~\ref{fig:ddp-unknown} for the case when mass hierarchy is known
and unknown, respectively. We see that if the mass hierarchy is known,
since $\delta^\prime$ = $\delta$ always holds when $\epsilon = 0$, the
diagonal line in the $\delta^\prime$ versus $\delta$ plot is always
allowed at less than $1\sigma$. If the mass hierarchy is unknown, when
the SM and NSI have the opposite mass hierarchy, there is a strong
correlation between $\delta$ and $\delta^\prime$ (which can be
described by $\delta^\prime = 180 -\delta$) as a result of the generalized MH
degeneracy. We also see that T2HKK has a better performance than T2HK
and DUNE in measuring $\delta$. In fact, if the mass hierarchy
is unknown, only T2HKK can measure $\delta$ at the $3\sigma$ CL
when three $\epsilon$'s are nonzero. 

\begin{figure}
\centering
\includegraphics[width=1.0\textwidth]{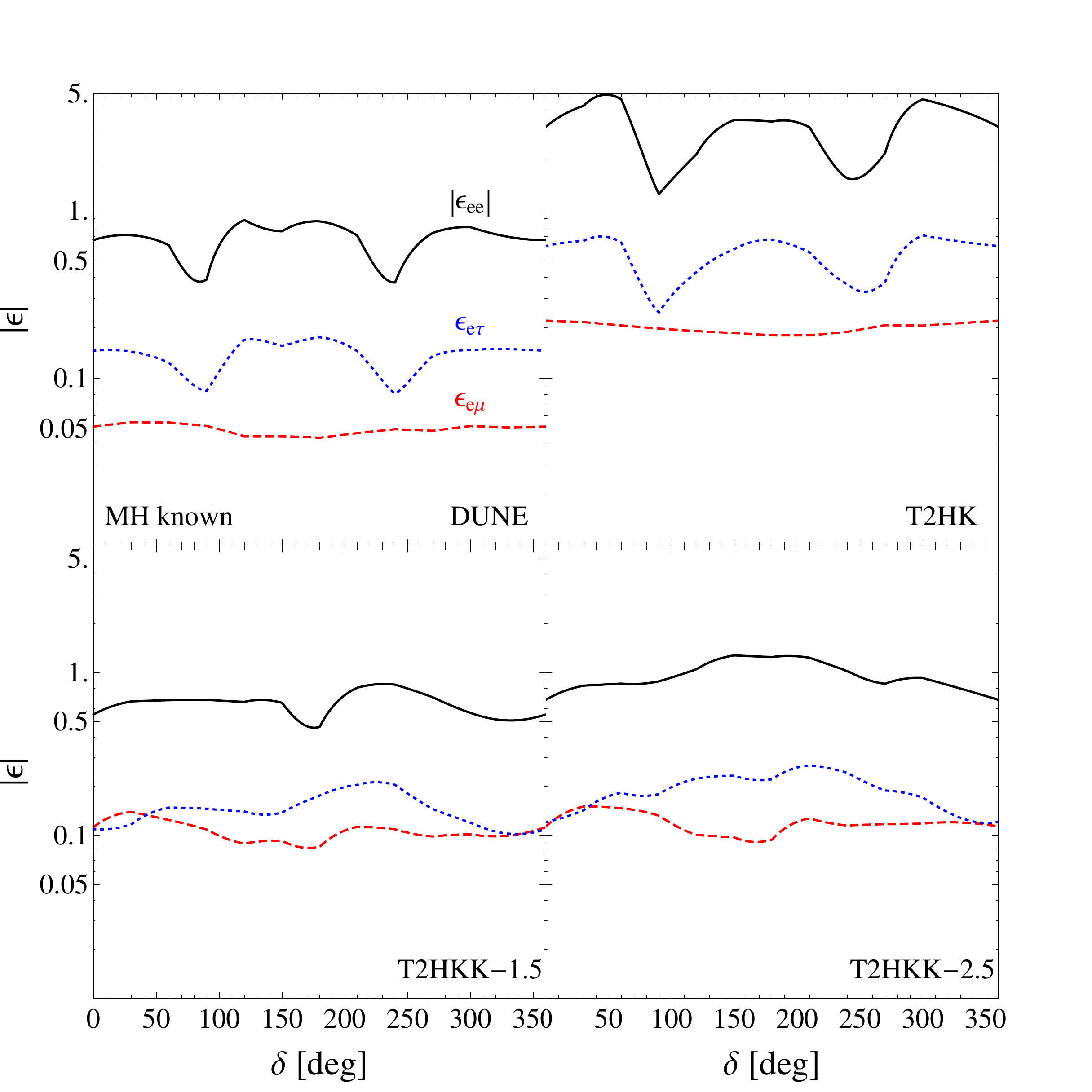}
\caption{The expected $2\sigma$ CL sensitivity to $|\epsilon|$ as a function of $\delta$ at DUNE, T2HK and T2HKK. The data are consistent with the SM and the NH. We assume only one $\epsilon$ is nonzero at a time and the mass hierarchy is known.
}
\label{fig:1eps-known}
\end{figure}

\begin{figure}
\centering
\includegraphics[width=1.0\textwidth]{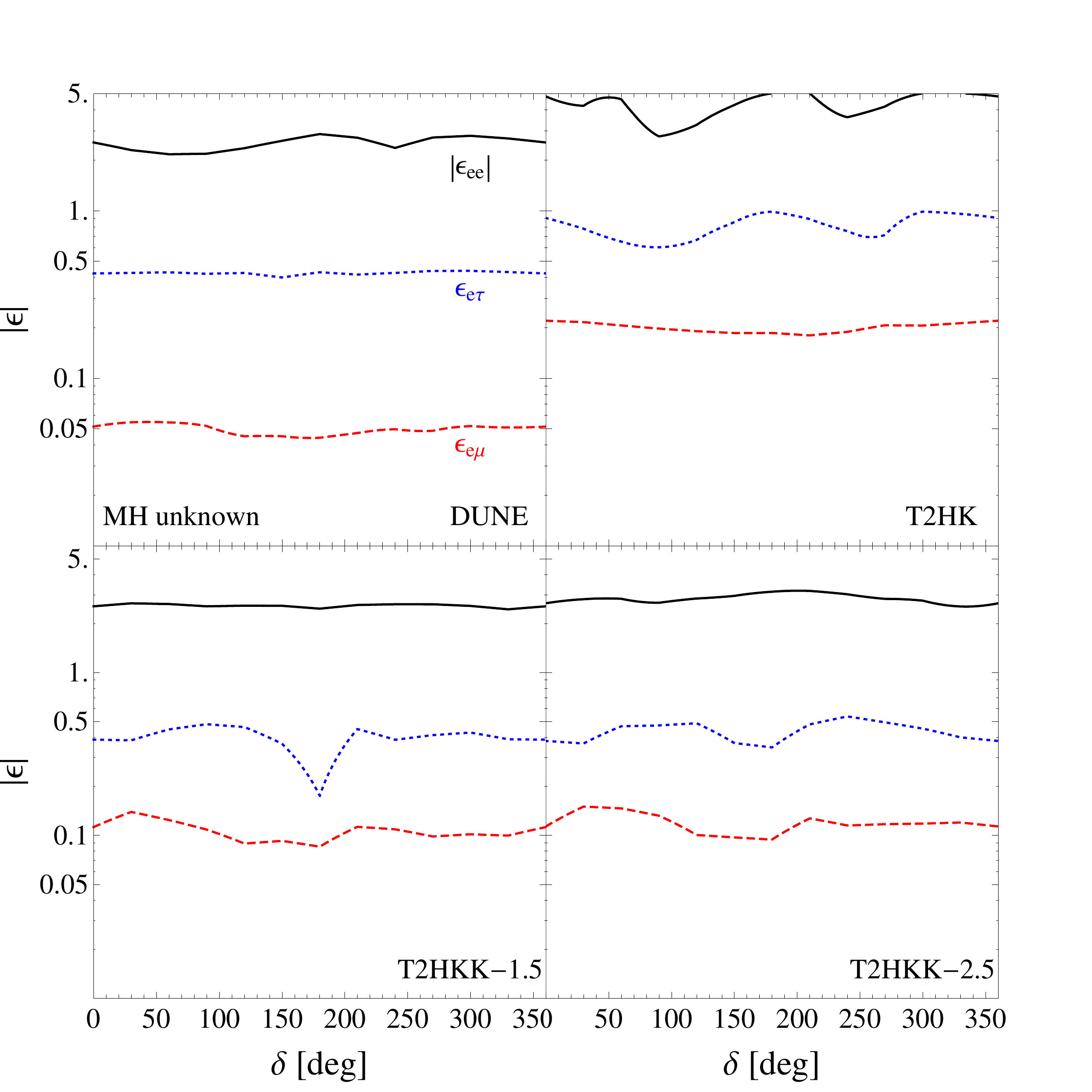}
\caption{Same as Fig.~\ref{fig:1eps-known}, except that the mass hierarchy is unknown.
}
\label{fig:1eps-unknown}
\end{figure}

We also plot the minimum value of $\epsilon$ for which the NSI scenario can be
discriminated from the SM at the 2$\sigma$ CL. If there is only one
nonzero $\epsilon$, the expected sensitivities are shown in
Figs.~\ref{fig:1eps-known} and~\ref{fig:1eps-unknown} if the MH is known
and unknown, respectively. As expected, the sensitivity is always weaker
if the MH is unknown than if the MH is known. Note that 
the minimum value of $|\epsilon_{ee}|$ that is detectable is always larger than 2 if the MH
is unknown due to the generalized MH degeneracy. From
Fig.~\ref{fig:1eps-known} we see that the sensitivity to
$|\epsilon_{ee}|$ and $|\epsilon_{e\tau}|$ at DUNE and T2HK improves
for $\delta \simeq 90$ and $240^\circ$, while this is not the case
for T2HKK. From Fig.~\ref{fig:1eps-unknown} we
see that there is a sharp improvement in the sensitivity to
$\epsilon_{e\tau}$ at T2HKK-1.5 for $\delta\simeq 180^\circ$ because
the IH is not allowed at the $2\sigma$ CL in this case.

\begin{figure}
\centering
\includegraphics[width=1.0\textwidth]{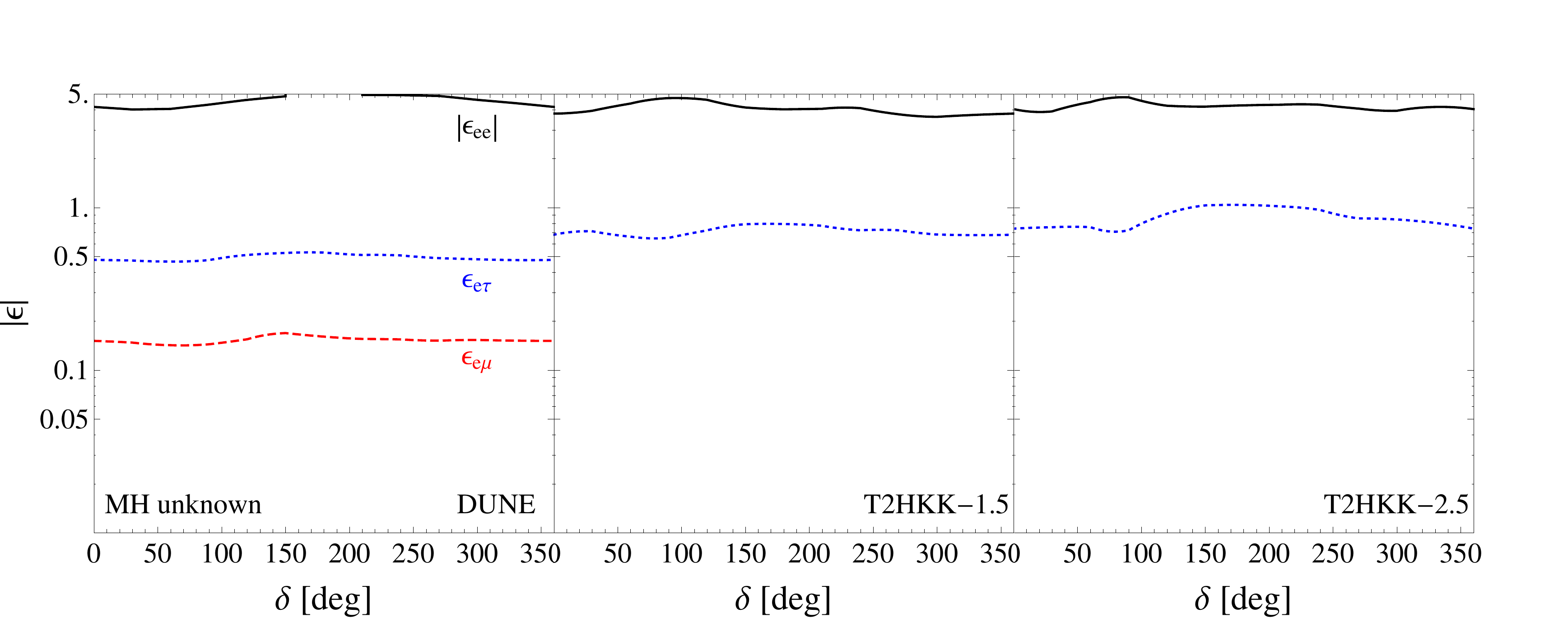}
\caption{The expected $2\sigma$ CL sensitivity to $|\epsilon|$ as a function of $\delta$ at DUNE and T2HKK. The data are consistent with the SM and the NH. We assume $\epsilon_{ee}$, $\epsilon_{e\mu}$ and $\epsilon_{e\tau}$ are all nonzero, and the mass hierarchy is unknown. 
For each curve, all the other parameters have been marginalized over.
All sensitivities for T2HK and the
$\epsilon_{e\mu}$ sensitivity for T2HKK are outside the scan range and are
therefore not shown. If the mass hierarchy is known, the sensitivities are unchanged for $\epsilon_{e\mu}$ and $\epsilon_{e\tau}$ and similar for $|\epsilon_{ee}|$.
}
\label{fig:3eps}
\end{figure}

In Fig.~\ref{fig:3eps} we show the expected sensitivities at 2$\sigma$
if $\epsilon_{ee}$, $\epsilon_{e\mu}$ and $\epsilon_{e\tau}$ are all
nonzero and the MH is unknown. We find that the sensitivities to all three
$\epsilon$'s at T2HK and the sensitivity to $\epsilon_{e\mu}$ at T2HKK
are outside the ranges that we scanned. Hence they are not shown
in Fig.~\ref{fig:3eps}. If
$\epsilon_{ee}$, $\epsilon_{e\mu}$ and $\epsilon_{e\tau}$ are all
nonzero, knowledge of the mass hierarchy does not affect the sensitivity to
$\epsilon_{e\mu}$ and $\epsilon_{e\tau}$ because we marginalize over the $\epsilon$'s thereby covering the regime of the generalized MH
degeneracy even if the MH is known. Furthermore, we see that the dependence on $\delta$ becomes
much weaker if all three $\epsilon$'s are nonzero, and that DUNE has
the best sensitivity to the magnitude of the NSI parameters overall. An examination of our figures shows that T2HKK-1.5 has better sensitivities than T2HKK-2.5 in both the SM and NSI scenarios.

\subsubsection{Sensitivity when the true mass hierarchy is inverted}

\begin{figure}
\centering
\includegraphics[width=0.8\textwidth]{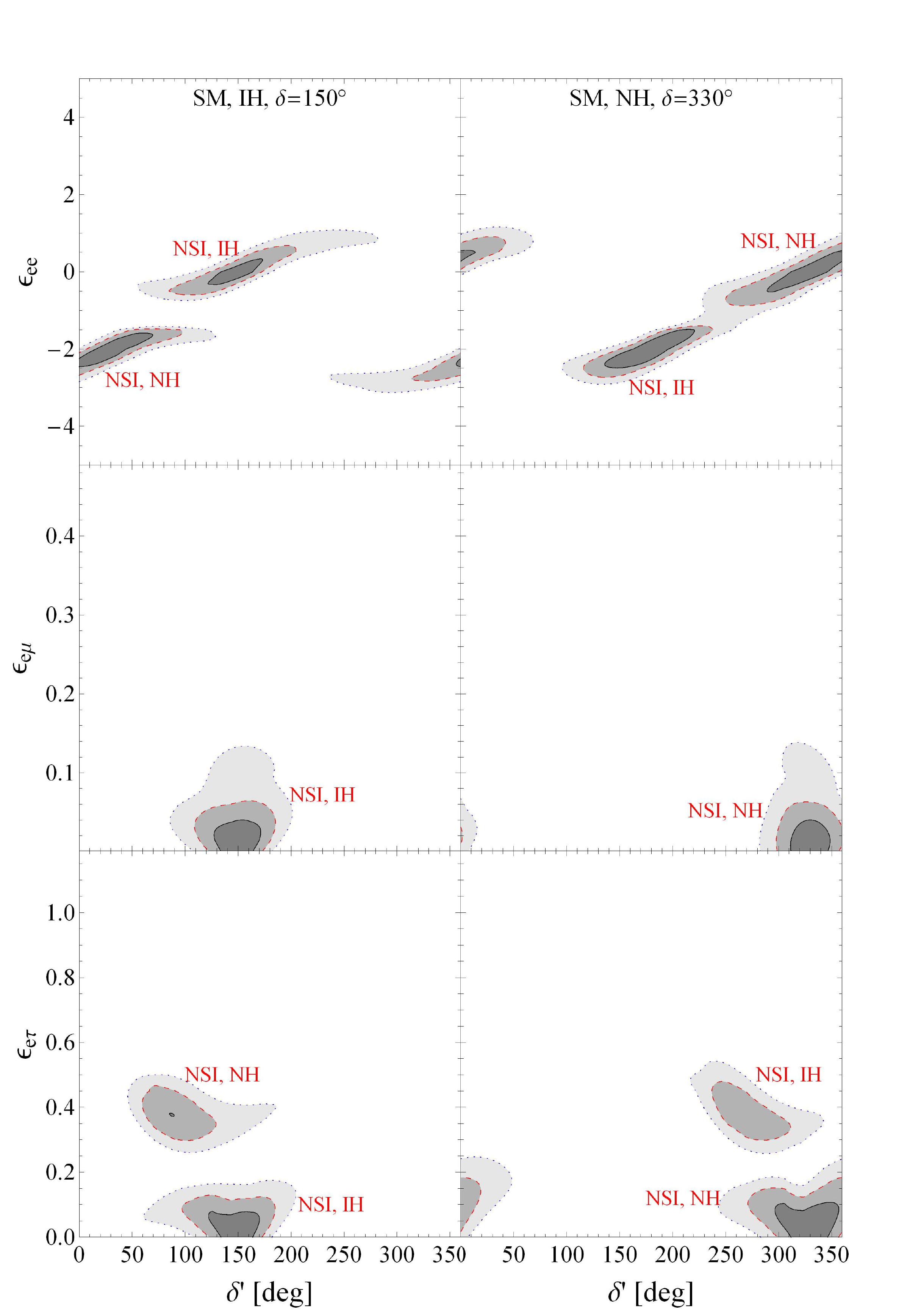}
\caption{$1\sigma$, $2\sigma$ and $3\sigma$ allowed regions for
  $\epsilon$ as a function of $\delta'$ at DUNE. The left (right)
  panels show the case when the data are consistent with the SM and the
  IH (NH) with $\delta=150^\circ$ ($330^\circ$). We fit the data
  assuming only one of $\epsilon_{ee}$, $\epsilon_{e\mu}$ or
  $\epsilon_{e\tau}$ is nonzero. We scan over both mass hierarchies
  and marginalize over the NSI phases.  }
\label{fig:duality}
\end{figure}

We now study the scenario in which the data are consistent with the SM
with the IH. We find that there is a similarity between the allowed
regions for when the data are consistent with the IH and the allowed regions
for when the data are consistent with the NH after a phase transformation in
$\delta$, i.e.,
\be
\delta \rightarrow \delta +180^\circ\,,\quad \delta'\rightarrow \delta'+180^\circ\,,\quad \phi_{\alpha\beta} \rightarrow \phi_{\alpha\beta} +180^\circ\,.
\label{eq:phasetrans}
\ee
An example of this similarity can be seen in Fig.~\ref{fig:duality},
in which we show the allowed regions for $\epsilon$ as a function of
$\delta'$ at DUNE for two cases: (a) the data are consistent with the
SM and the NH with $\delta=330^\circ$, and (b) the data are consistent
with the SM and the IH with $\delta=150^\circ$. We see that
a shift of $\delta'\rightarrow \delta'+180^\circ$, renders the allowed regions
between the two cases very similar; the
allowed regions of the NSI scenario with the IH (NH) in case (a) are
similar to the allowed regions of the NSI scenario with the NH (IH) in
case (b) after the phase transformation.

This similarity can be understood as follows. In order to fit the SM
data with an NSI scenario, the two main constraints from the
appearance channel in both the neutrino and antineutrino modes are
\bea
&P^\text{NSI}_i(\nu_\mu\rightarrow\nu_e)- P^\text{SM}_j(\nu_\mu\rightarrow\nu_e)=0\,,
\\
&P^\text{NSI}_i(\bar{\nu}_\mu\rightarrow\bar{\nu}_e)-P^\text{SM}_j(\bar{\nu}_\mu\rightarrow\bar{\nu}_e)=0\,,
\eea
where $i,j= \text{NH},\text{IH}$.

Using Eq.~(\ref{eq:app-prob}) at leading order in $\epsilon$,  we have
\bea
0&=&[x^2 f^2 + 2xyfg \cos(\Delta + \delta')+y^2g^2] - [x^2 \bar{f}^2 - 2xy\bar{f}g \cos(-\Delta + \delta)+y^2g^2]
\nonumber\\
&+&4\hat A \epsilon_{e\mu} x [s_{23}^2 f^2 \cos(\phi_{e\mu}+\delta')  
+ c_{23}^2  f g \cos(\Delta+\delta'+\phi_{e\mu})]
\nonumber\\
&+& 4\hat A \epsilon_{e\tau} x s_{23} c_{23}
  [f^2 \cos(\phi_{e\tau}+\delta')  
- f g \cos(\Delta+\delta'+\phi_{e\tau})] \,,
\label{eq:nu}
\\
0&=&[x^2 \bar{f}^2 + 2xy\bar{f}g \cos(\Delta - \delta')+y^2g^2] - [x^2 f^2 - 2xyfg \cos(-\Delta - \delta)+y^2g^2]
\nonumber\\
&-&4\hat A \epsilon_{e\mu} x [s_{23}^2 \bar{f}^2 \cos(\phi_{e\mu}+\delta')  
+ c_{23}^2  \bar{f} g \cos(\Delta-\delta'-\phi_{e\mu})]
\nonumber\\
&-& 4\hat A \epsilon_{e\tau} x s_{23} c_{23}
  [\bar{f}^2 \cos(\phi_{e\tau}+\delta')  
- \bar{f} g \cos(\Delta-\delta'-\phi_{e\tau})] \,,
\label{eq:antinu}
\eea
for the NSI scenario with the NH and the SM scenario with the IH. 
 Switching the mass hierarchy of the SM
and NSI scenarios  (via $\Delta \to - \Delta$, $f\leftrightarrow - \bar{f}$, and $g \to -g$),
and applying the phase transformation of
Eq.~(\ref{eq:phasetrans}), leads to an interchange of
Eq.~(\ref{eq:nu}) and Eq.~(\ref{eq:antinu}) so that we obtain the same two
constraints on the NSI parameters. Since the phase transformation does not
depend on $L$ and $E$, the allowed regions at DUNE, T2HK
and T2HKK are similar for both hierarchies. However, note that if we take into account the higher
order terms in Eq.~(\ref{eq:app-prob}), in particular the third and
fifth lines in Eq.~(\ref{eq:app-prob}), the phase transformation does not leave the two constraints unchanged, which explains the small difference between the allowed regions in
the two scenarios. There is a similar correspondence for any
combination of hierarchies between the SM and NSI scenarios.

\begin{figure}
\centering
\includegraphics[width=1.0\textwidth]{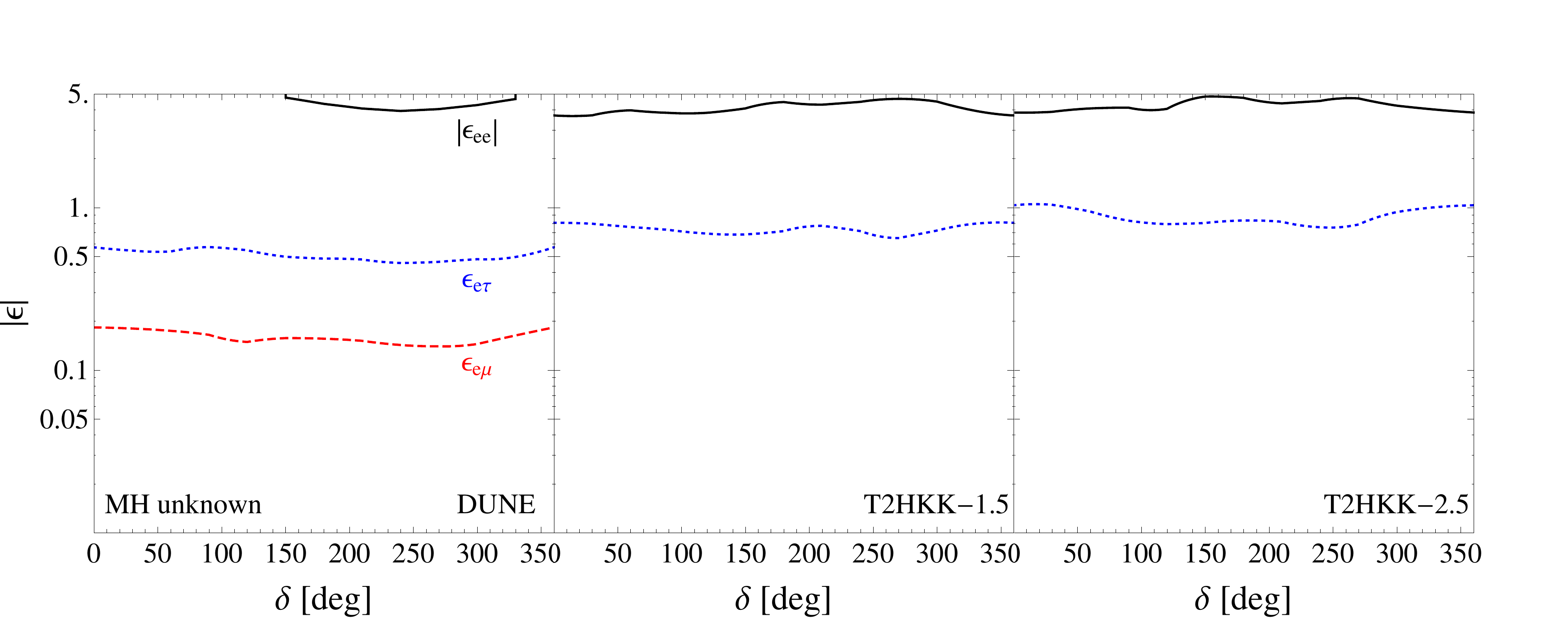}
\caption{Same as Fig.~\ref{fig:3eps}, except that the data are consistent with the SM and the IH. }
\label{fig:3eps-smih}
\end{figure}

Because of the correspondence discussed above, we
expect the NSI sensitivities to be similar whether the data are
consistent with the SM in the NH or the SM in the IH. This can seen by
comparing Fig.~\ref{fig:3eps} with Fig.~\ref{fig:3eps-smih}.
In sum, the NSI sensitivities
in a given experiment will be similar regardless of the true hierarchy, modulo the transformation $\delta \to \delta +180^\circ$.

\subsection{NSI in the disappearance channels ($\epsilon_{\mu\tau}$ and $\epsilon_{\tau\tau}$)} 

\begin{figure}
\centering
\includegraphics[width=0.8\textwidth]{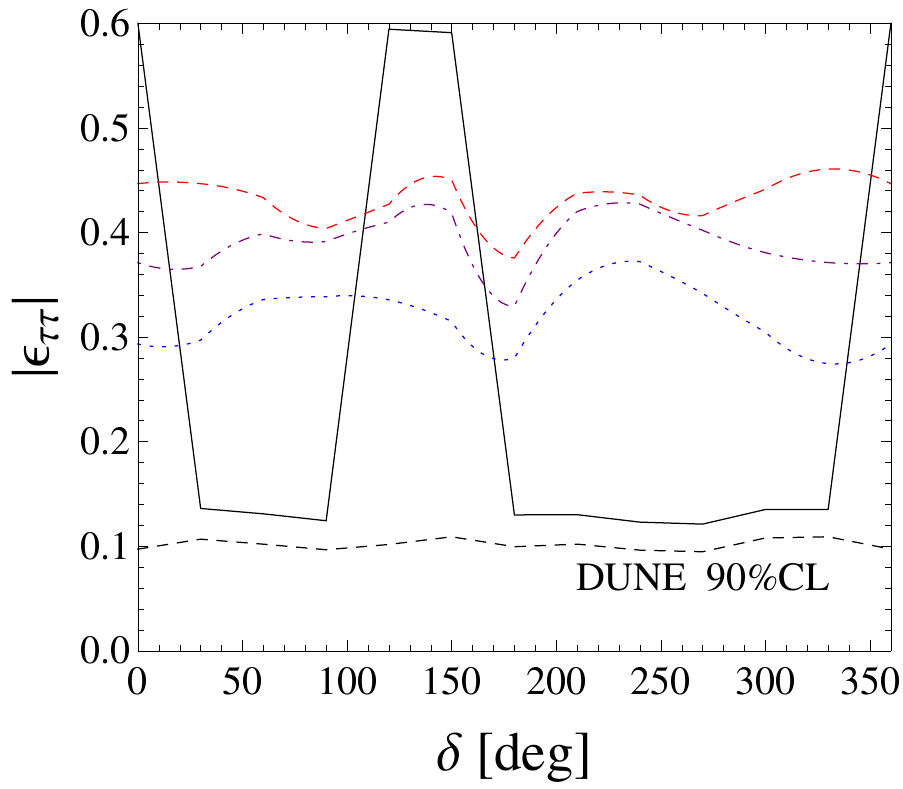}
\caption{The expected $2\sigma$ sensitivity to $|\epsilon_{\tau\tau}|$
  as a function of $\delta$.  The black solid, red dashed, blue
    dotted, and purple dotdashed curves correspond to DUNE, T2HK,
  T2HKK-1.5, and T2HKK-2.5, respectively. The data are consistent with the
  SM and the NH. We assume only $\epsilon_{\mu\tau}$ and
  $\epsilon_{\tau\tau}$ are nonzero, and all the parameters not
  shown have been marginalized over. We also show the 90\%~CL
  sensitivity curve for DUNE to emphasize that the sensitivity is uniform if degenerate regions
  close to the boundaries of the scanned NSI parameter range are excluded by atmospheric data; see Fig.~\ref{fig:tt-delptap} for an illustration of the degenerate region.  }
\label{fig:tt-2eps}
\end{figure}

\begin{figure}
\centering
\includegraphics[width=0.8\textwidth]{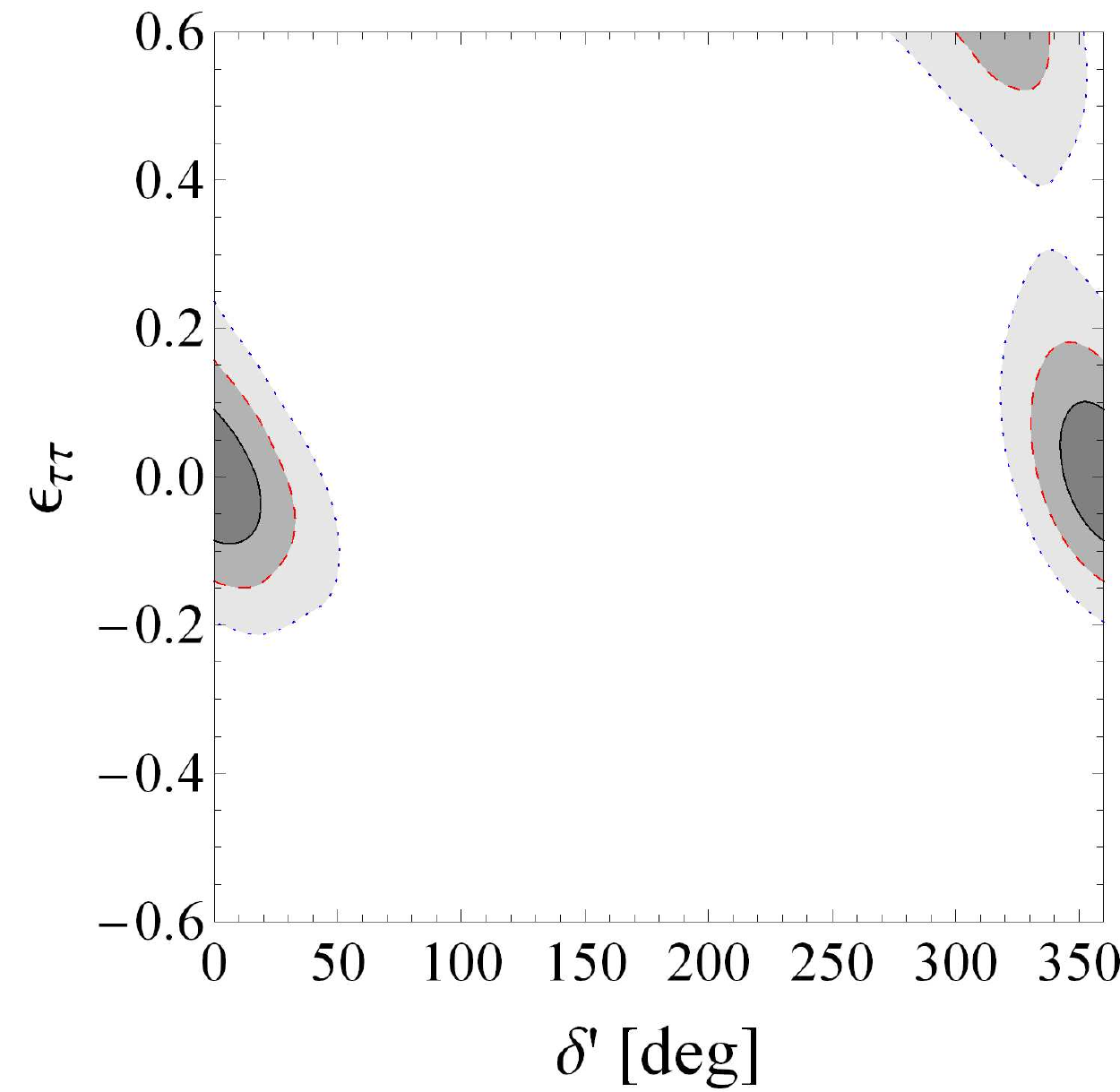}
\caption{$1\sigma$, $2\sigma$ and $3\sigma$ allowed regions in the
  $\epsilon_{\tau\tau}$ versus $\delta'$ plane at DUNE. The data are
  consistent with the SM with $\delta = 0$ and the NH. We assume that
  only $\epsilon_{\mu\tau}$ and $\epsilon_{\tau\tau}$ are nonzero and
  all the parameters not shown have been marginalized over.  }
\label{fig:tt-delptap}
\end{figure}

We find that the sensitivity of these experiments to
$\epsilon_{\mu\tau}$ is outside the range of our scan. After marginalizing over
$\epsilon_{\mu\tau}$ and the mass hierarchy, we show the $2\sigma$
sensitivities to $\epsilon_{\tau\tau}$ at DUNE, T2HK and T2HKK in
Fig.~\ref{fig:tt-2eps}. We see that T2HKK-1.5 has better sensitivity
than T2HKK-2.5 because of its higher energy spectrum and larger statistics.  We also see
that the $2\sigma$ sensitivity at DUNE becomes quite weak at some
$\delta$ values. This is due to a degenerate region near the boundary
of the ranges that we have scanned, which was first noticed in
Ref.~\cite{Coloma:2015kiu} and can be seen in
Fig.~\ref{fig:tt-delptap}. Since this degenerate region, which occurs
because of a correlation between $\epsilon_{\tau\tau}$ and the
deviation of $\theta_{23}$ from maximal mixing, is at the
boundary, we also
show the 90\% CL sensitivity curve at DUNE in Fig.~\ref{fig:tt-2eps}
to emphasize that the sensitivity is uniform if the degenerate region
is resolved by future atmospheric data.

\section{Summary}

We studied the sensitivities to NSI in the proposed next generation
long-baseline neutrino experiments DUNE, T2HK and T2HKK. For the T2HK
and T2HKK experiments, we adopted the new detector configurations and fluxes
provided by the Hyper-Kamiokande collaboration.

To understand the effect of each NSI parameter on the
experimental performance, we considered different scenarios with
different combinations of NSI parameters. We find that if only one of
$\epsilon_{e\mu}$ or $\epsilon_{e\tau}$ is nonzero, most of the
continuous four-fold degeneracies with NSI at a single $L$ and $E$
measurement can be resolved for the range of energies available at
these experiments. The degeneracies are broken even further at T2HKK
with two different baselines. However, if multiple NSI are nonzero,
these experiments can not measure the mass hierarchy, CP phase and
$\theta_{23}$ octant as a result of degeneracies between NSI and SM
parameters, and between NSI parameters. As a specific realization of the latter, 
we find that a cancellation between terms
at leading order in the appearance channel probabilities when
$\epsilon_{e\tau} = \cot\theta_{23}\epsilon_{e\mu}$ strongly
affects the sensitivities to these two NSI parameters at T2HK and
T2HKK.  Also, the sensitivities at all three experiments are worsened by
the generalized mass hierarchy degeneracy in the NSI scenario. Because
the generalized mass hierarchy degeneracy occurs at the Hamiltonian
level, atmospheric neutrino and reactor neutrino experiments will not be able to resolve it.

We also studied the dependence of the sensitivities on the true CP phase
$\delta$ and the true mass hierarchy. We find that the sensitivities are
much weaker for all values of $\delta$ when multiple NSI are
nonzero. Also, we find that due to leading order effects in the
appearance channel probabilities, there is a similarity of the allowed
regions for the NSI parameters between the case in which the data are
consistent with the normal hierarchy and the case in which the data are
consistent with the inverted hierarchy. Thus the sensitivities
are similar whether nature has chosen the NH or the IH, modulo the
transformation $\delta \rightarrow \delta+180^\circ$. 

Overall DUNE has the best sensitivity to the magnitude of the NSI parameters,
while T2HKK has the best sensitivity to CP violation whether or not there are NSI, and
overall T2HKK-1.5 does better than T2HKK-2.5.

We further studied the sensitivities to $\epsilon_{\mu\tau}$ and
$\epsilon_{\tau\tau}$ that mainly come from the disappearance
channel. We find that the sensitivities to $\epsilon_{\mu\tau}$ are
limited compared to atmospheric experiments, and we obtained the
sensitivity to $\epsilon_{\tau\tau}$ at these three experiments.

~\\
{\bf Acknowledgments.}

We thank S.-H. Seo and M. Yokoyama for useful inputs regarding T2HK and T2HKK.
KW thanks the University of Hawaii at Manoa for its hospitality during
part of this work. This research was supported by the U.S. DOE under
Grant No. DE-SC0010504.

\clearpage

%
%
%

\clearpage

\end{document}